\documentclass{article}

\usepackage{microtype}
\usepackage{graphicx}
\usepackage{booktabs} 
\usepackage{hyperref}

\usepackage{listings}
\usepackage{graphicx}
\usepackage{xcolor}
\usepackage{xspace}
\usepackage{subcaption}
\usepackage{enumitem}
\usepackage{flushend}
\hypersetup{draft}

\newcommand{\parab}[1]{\vspace{0.05in}\noindent\textbf{#1}}

\newcommand{\sysname}{\textsc{Caramel}\xspace}

\usepackage[accepted]{sysml2019}

\sysmltitlerunning{\sysname: Accelerating Decentralized Distributed Deep Learning with Model-Aware Scheduling}

\begin{document}

\twocolumn[
\sysmltitle{\sysname: Accelerating Decentralized Distributed Deep Learning with Computation Scheduling}




\sysmlsetsymbol{equal}{*}

\begin{sysmlauthorlist}
\sysmlauthor{Sayed Hadi Hashemi}{equal,goog}
\sysmlauthor{Sangeetha Abdu Jyothi}{equal,uci,vmwr}
\sysmlauthor{Brighten Godfrey}{uiuc,vmw}
\sysmlauthor{Roy Campbell}{uiuc}
\end{sysmlauthorlist}

\sysmlaffiliation{goog}{Google}
\sysmlaffiliation{uci}{UC Irvine}
\sysmlaffiliation{vmwr}{VMware Research}
\sysmlaffiliation{vmw}{VMware}
\sysmlaffiliation{uiuc}{UIUC}

\sysmlcorrespondingauthor{Sayed Hadi Hashemi}{hashemi3@illinois.edu}
\sysmlcorrespondingauthor{Sangeetha Abdu Jyothi}{sangeetha.aj@uci.edu}


\vskip 0.3in

\begin{abstract}
The method of choice for parameter aggregation in Deep Neural Network (DNN) training, a network-intensive task, is shifting from the Parameter Server model to decentralized aggregation schemes (AllReduce) inspired by theoretical guarantees of better performance. However, current implementations of AllReduce overlook the interdependence of communication and computation, resulting in significant performance degradation. In this paper, we develop Caramel, a system that accelerates decentralized distributed deep learning through model-aware computation scheduling and communication optimizations for AllReduce. Caramel achieves this goal through (a) computation DAG scheduling that expands the feasible window of transfer for each parameter (\textit{transfer boundaries}), and (b) network optimizations for smoothening of the load including adaptive batching and pipelining of parameter transfers. Caramel maintains the correctness of the dataflow model, is hardware-independent, and does not require any user-level or framework-level changes. We implement Caramel over TensorFlow and show that the iteration time of DNN training can be improved by up to $3.62\times$ in a cloud environment.
\end{abstract}
]
\printAffiliationsAndNotice{\sysmlEqualContribution} 

\section{Introduction}

Deep Neural Networks (DNNs) form the crux of advanced solutions in a variety of fields such as computer vision and natural language processing. In frameworks such as TensorFlow~\cite{abadi2016tensorflow}, the interdependence of computation and communication operations involved in training a model is represented using a dataflow graph, which is a Directed Acyclic Graph (DAG). The state of the DNN is represented by a vector of parameters. Each iteration involves the computation of parameter updates, followed by its exchange between the participating nodes.

Today, performance and scalability of distributed DNN training in the cloud are bottlenecked by this parameter aggregation~\cite{FireCaffe, dnnLimits}. Recently, decentralized aggregation schemes~\cite{barnett1994interprocessor, thakur2003improving, Dean:2008:MSD:1327452.1327492} have emerged as a popular choice of aggregation in many frameworks~\cite{baidu-allreduce, horovod, tfdistri87:online}. In these schemes, unlike in the Parameter Server model, parameters are aggregated through collective transfers such as \texttt{MPI\_allreduce()} in MPI~\cite{MPI} and \texttt{ncclAllReduce()} in Nvidia's NCCL. However, in spite of recent optimizations~\cite{goyal2017accurate}, current decentralized implementations fail to achieve the guaranteed performance gains since they overlook interdependency of communication and computation, especially in the cloud, leaving GPUs idle for a significant fraction of time. 

In this paper, we introduce \sysname to improve efficiency of decentralized DNN training, in terms of iteration time and GPU utilization, through model-aware dataflow DAG optimizations. \sysname achieves this goal through (a) computation scheduling that expands the feasible window of transfer for each parameter (\textit{transfer boundaries}) and (b) network optimizations that smoothen the load. 

The transfer boundaries of a parameter represent the window when that parameter can be aggregated without blocking computation. When the transfer boundaries are farther apart, the performance is less affected by a slow network. \sysname expands these boundaries through scheduling optimizations of the computation DAG where it (i) moves the start boundaries earlier while also reducing variance and (ii) pushes the end boundary by postponing the execution of some computation operations to the forward pass of next iteration. Optimizations for smoothening the network load include (iii) batching of small parameters to reduce the network overhead, and (iv) adaptive splitting and pipelining of parameters to accelerate aggregation of large data which involves multi-stage network transfers with intermediate aggregation computation at workers. 

Optimizations in \sysname are motivated by following observations of shortcomings in state-of-the-art decentralized aggregation systems.

First, a dataflow model (DAG) may have multiple feasible traversals, i.e., different orders of execution for computation operations in the DAG which are all valid. Network transfers are leaf nodes in this dataflow DAG. Based on the schedule chosen for computation operations, network transfers may be activated (i.e., parameters being ready for aggregation) in different orders across multiple workers. This can prove detrimental in decentralized aggregation where all workers should activate the same parameter before its transfer can be initiated and bad schedules can delay transfers. To solve this problem, \sysname enforces a schedule on network transfers by adding additional dependencies in the DAG to force all workers to follow the same best schedule. Note that this does not affect the correctness of the initial model as \sysname is enforcing one of the valid schedules in the initial DAG, thereby reducing the variance in start boundary of each parameter.

Second, we identify an opportunity for increasing the window of network transfer during an iteration by pushing the end boundary. An iteration has two phases: forward pass and backpropagation phase. Currently, transfers are restricted to the backpropagation phase. We propose techniques for extending network transfers to forward pass in \sysname by postponing execution of some network operations to the forward pass of the next iteration in the dataflow DAG without affecting computation operations.

Third, all DNNs we analyzed have a large number of small parameters which incur significant overhead during network transfer. To tackle the small-parameter overhead, we implement model-aware batching in \sysname, while also ensuring that the batched parameters are ready at nearly the same time to avoid waiting.

Fourth, transfer of large parameters can be accelerated by splitting a single large aggregation operation into multiple smaller aggregations over partitions of the data and pipelining computation and communication stages of each sub-operation. \sysname adaptively chooses the optimal level of splitting them.

To the best of our knowledge, \sysname is the first work that improves overlap of communication and computation in DNN training solely by \textit{computation} scheduling/transformation. The network optimizations in \sysname (adaptive batching and splitting) are also implemented as DAG operations. \sysname advances state of the art by generating an optimized DAG with several practical benefits: (a) \sysname can use all accelerators/network fabrics supported by ML framework out of the box, (b) \sysname is portable and compatible with ML pipeline services such as fault recovery/checkpointing, and (c) no modifications to ML frameworks or external dependencies are required with \sysname.

We implement \sysname over TensorFlow and demonstrate that the iteration time can be reduced by up to $3.62\times$, with up to $73\%$ network cost reduction. In summary, we make the following contributions:
\vspace{-2mm}
\begin{itemize}
    \itemsep0em
    \item We identify opportunities for improving efficiency of decentralized distributed DNN training.
    \item We develop \sysname and implement it over TensorFlow with model-aware optimizations to expand \textit{transfer boundaries} and smoothen network utilization.
    \item We extensively evaluate the performance of \sysname in the Azure cloud and show that training iteration time can be improved by up to $3.62\times$ and GPU utilization by up to $3\times$ in $5$ commonly used DNN models. 
\end{itemize}

\section{Background}
In this section, we give a brief overview of the distributed DNN training environment that we aim to optimize.

\begin{figure}
\centering
\begin{subfigure}[h]{\columnwidth}
  \centering
  \includegraphics[width=0.8\textwidth]{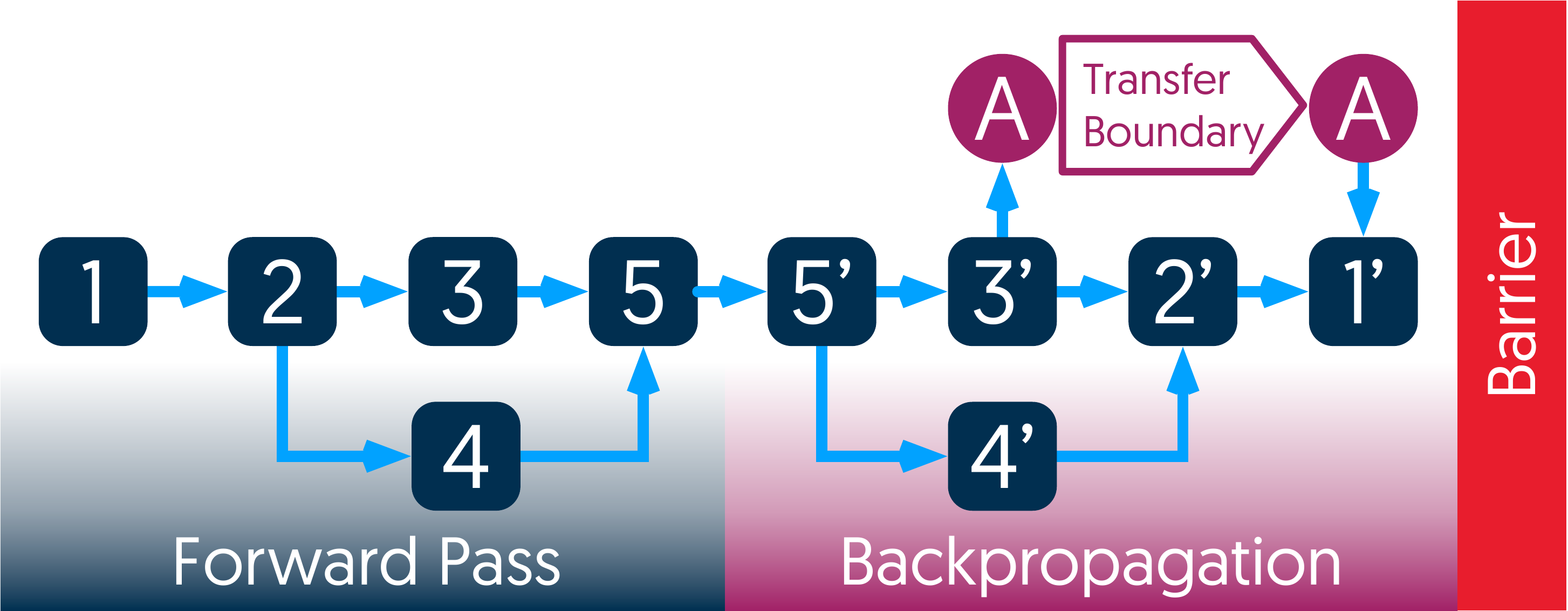}
  \caption{Example DAG}
  \label{fig:example_dag}
  \vspace{2mm}  
\end{subfigure}
\begin{subfigure}[t]{\columnwidth}
  \includegraphics[height=0.95cm]{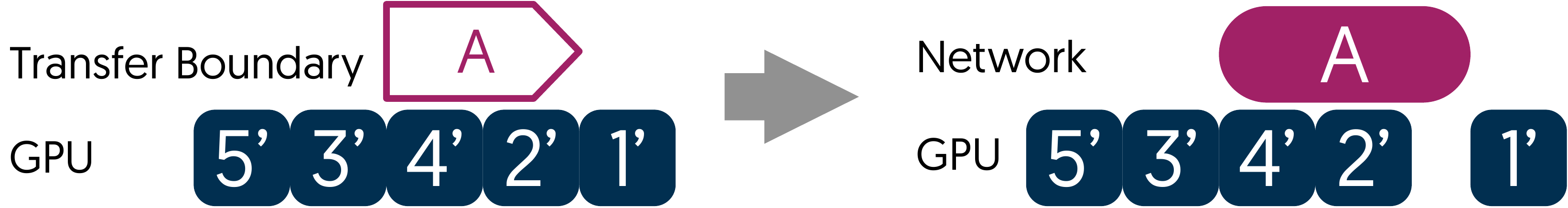}
  \caption{Best Schedule}
  \label{fig:example_dag_best}
  \vspace{2mm}
\end{subfigure}
\begin{subfigure}[t]{\columnwidth}
  \includegraphics[height=0.95cm]{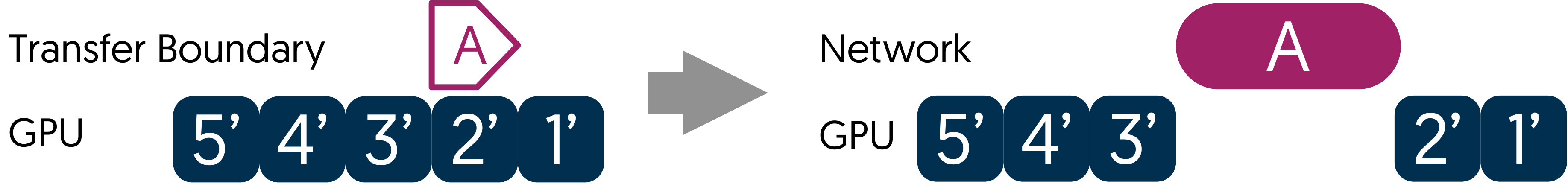}
  \caption{Worst Schedule}
  \label{fig:example_dag_worst}
\end{subfigure}

\caption{Impact of Transfer Window on Performance}
\label{fig:transfer_window_example}
\vspace{-2mm}
\end{figure}

Popular machine learning frameworks such as TensorFlow~\cite{abadi2016tensorflow} and PyTorch~\cite{paszke2017pytorch} represent the DNN training as a Directed Acyclic Graph (DAG). A toy model is given in Figure~\ref{fig:example_dag}. A DAG has two types of operations (ops): computation ops (multiplication, convolution etc.) and communication ops (read and update). Each parameter is read and updated independently. Each iteration has two stages: forward pass and backpropagation phase. In the forward pass, a loss function is calculated based on the input to the model. In the backpropagation phase, the model parameters are updated based on the calculated loss.

We target the commonly used model replica (MR) (also called data parallel) style of distributed training. In this style, each participating node called worker has a copy of the complete DAG. Input data is partitioned and fed in parallel to the workers. A worker computes updates (gradients) to model parameters based on its inputs. Update to a given parameter is of the same size (byte length) at all workers and the aggregation process is typically a commutative operation (mainly addition). In synchronized training in model replica there is a barrier at the end of iteration to ensure all the workers have their updates aggregated.

Parameter aggregation can be done in several ways. In Parameter Server (PS) mode, there are one or more centralized servers responsible for aggregating parameters. In this paper, our focus is on \textbf{decentralized aggregation} techniques (Bucket or Ring algorithm~\cite{barnett1994interprocessor}, Vector Halving and Distance Doubling Algorithm (HD)~\cite{thakur2003improving}, Shuffle \cite{Dean:2008:MSD:1327452.1327492}, etc.). In all decentralized patterns, aggregation of a parameter is initiated only after it is activated at all workers. Unlike PS, all workers are involved in the process with communication and computation load related to aggregation distributed across nodes based on the pattern selected. Currently, decentralized aggregation is initiated for each parameter in the backpropagation phase after the parameter is updated.

\section{Motivation}
\label{sec: Motivation}
In distributed DNN training, GPUs are forced to be idle when waiting for network transfers to complete. In this section, we define transfer boundaries of a parameter and analyze various factors causing delays in DNN training. 

\begin{figure}
\centering
\includegraphics[width=\columnwidth]{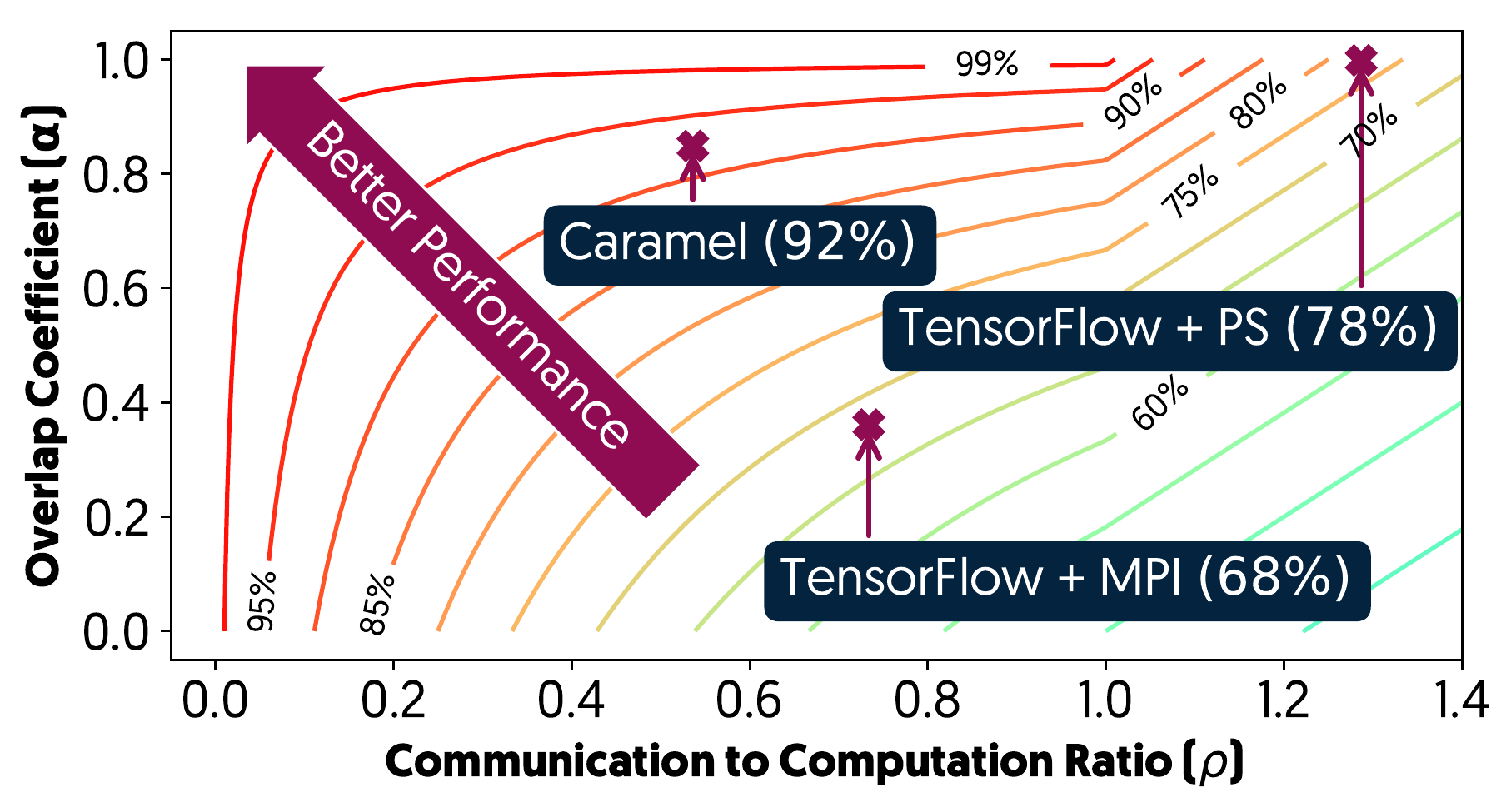}%
\caption{Overlap coefficient and communication/computation ratio for different frameworks with GPU utilization contours in the background (using Inception-v3 with $8$ workers).}
\label{fig:inception_example}
\vspace{-5mm}
\end{figure}


\subsection{Defining the Environment}
The total iteration time ($T$), communication time ($N$), and the computation time ($C$) are related as $T \leq N + C$ since the computation and communication may overlap. As shown in \cite{tictac}, the communication/computation ratio, $\rho$, the overlap coefficient, $\alpha$, and the GPU utilization, $U$, are related as follows: $U = \frac{1}{1+\rho - \alpha * min(\rho,1)}$. When $\rho < 1$, communication time is smaller than the total computation time, providing ample opportunity for running GPUs at high utilization. Poor overlap of communication and computation can result in low GPU utilization.

\subsection{Performance of current systems}
Similar to PS comparisons in \cite{tictac}, we plot the contour curves for GPU utilization with respect to $\alpha$ and $\rho$ to understand the performance of MPI implementation in the state-of-the-art decentralized aggregation with Horovod~\cite{horovod}. We observe that TensorFlow with Horovod suffers from poor overlap of communication and computation, and hence poor GPU utilization. In this paper, we will identify causes for this poor performance and design optimizations in \sysname that help us improve GPU utilization significantly. 

\begin{figure}
\centering
\begin{subfigure}[h]{0.49\columnwidth}
  \centering
  \includegraphics[width=\textwidth]{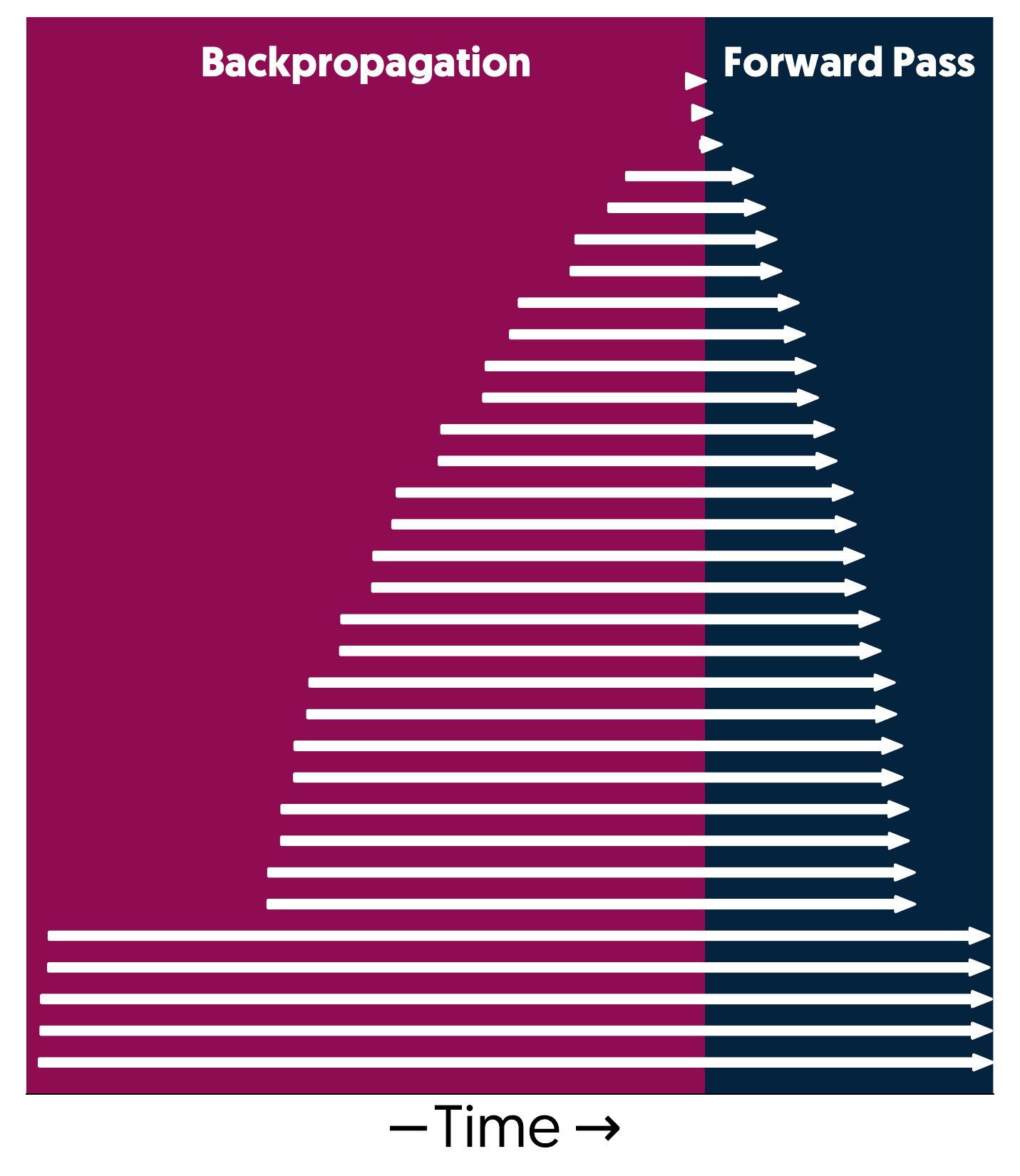}%
  \label{fig:transfer_window_ps}
  \caption{Parameter Server}
\end{subfigure}
\begin{subfigure}[h]{0.49\columnwidth}
  \centering
  \includegraphics[width=\textwidth]{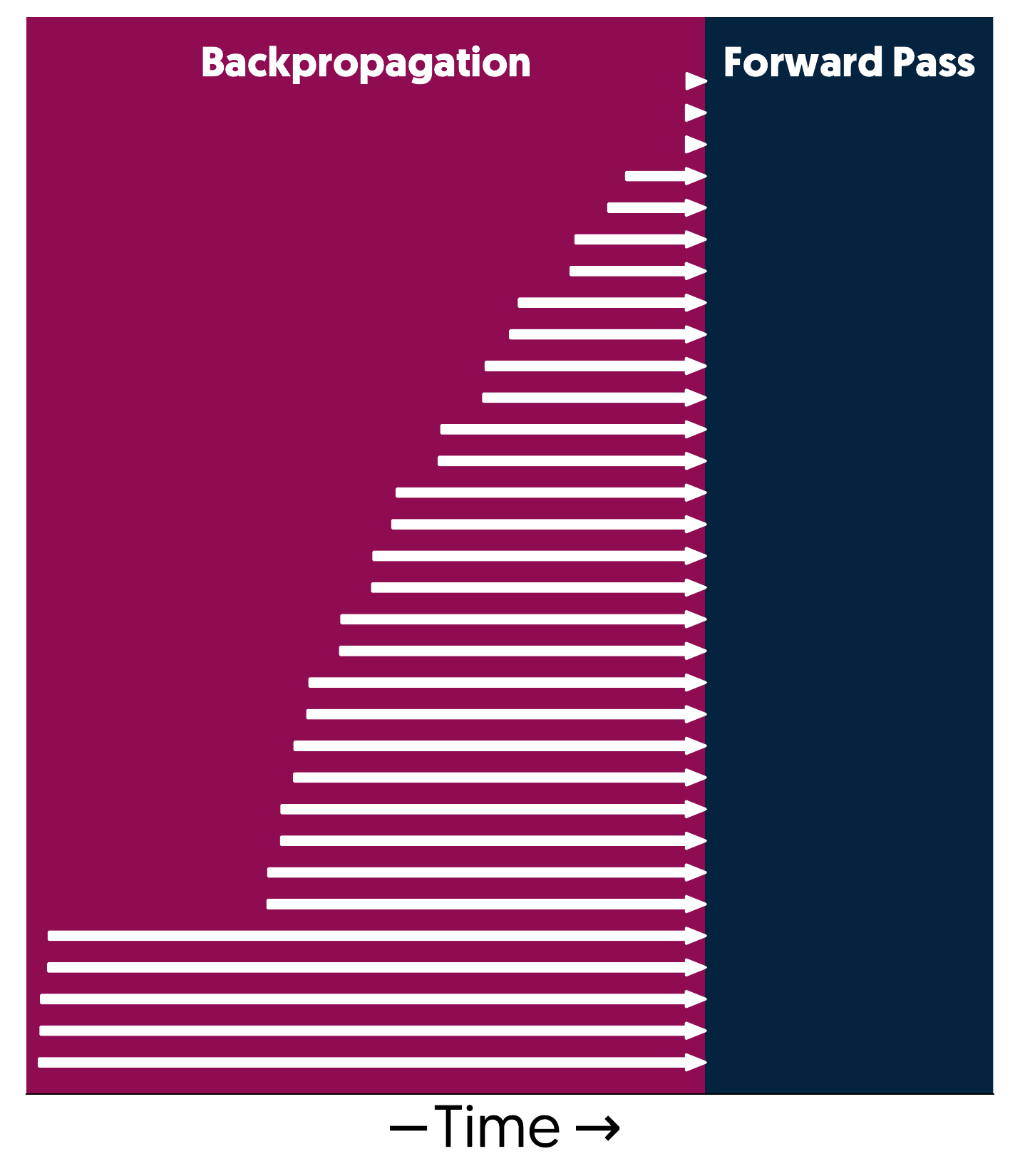}%
  \caption{Decentralized aggregation}
  \label{fig:transfer_window_ar}
\end{subfigure}
\caption{Comparison of transfer boundaries in a single iteration of distributed training with PS and decentralized aggregation. Data collected from training VGG-16 with batch size of 256.}
\label{fig:transfer_window_comparision}
\vspace{-2mm}
\end{figure}








\subsection{Understanding Model Characteristics}
\label{sec:modelChar}
Next, we define transfer window of a parameter and identify causes of low GPU utilization based on this characteristic and other well-known attributes of a model.

\subsubsection{Transfer Boundary}
We define \textbf{transfer boundary} of a parameter based on the window where its aggregation is feasible. The start boundary is determined by the completion of the computation operation that updates the parameter. The end boundary is the computation operation that reads the parameter. Given a schedule of computation operations, start and end boundaries of a parameter are fixed. For example, in Figure~\ref{fig:example_dag}, start boundary is at $3'$ where parameter $A$ is updated and end boundary is at $1'$ where parameter $A$ is read.

\begin{figure*}
\centering
\begin{subfigure}[t]{0.30\textwidth}
\centering
    \includegraphics[width=\textwidth]{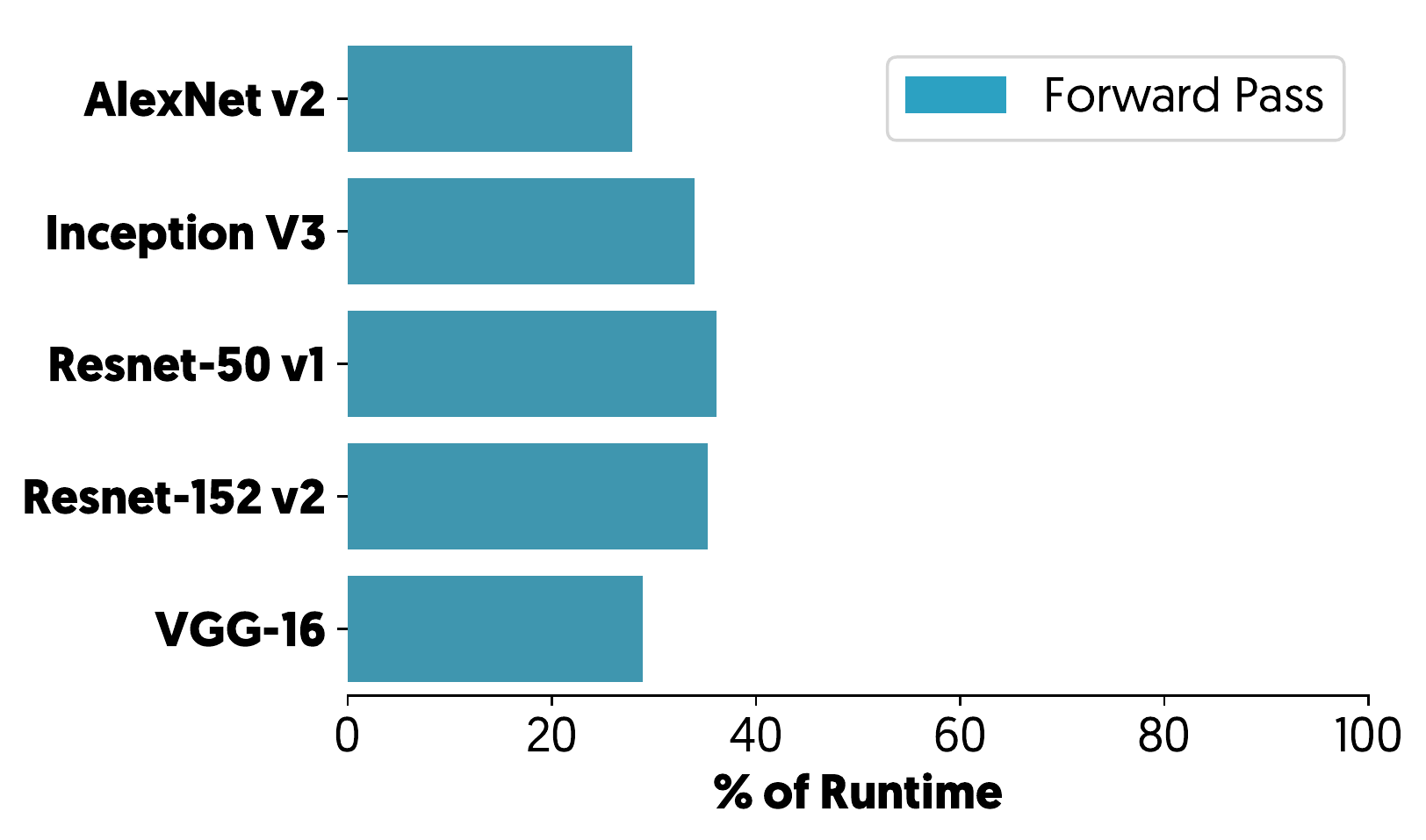}%
\label{fig:fp-bp}
\caption{\small Percentage of Forward Pass in common DNN models}
\end{subfigure}
\hspace{0.01\textwidth}
\begin{subfigure}[t]{0.38\textwidth}
\centering
 \includegraphics[width=0.9\textwidth]{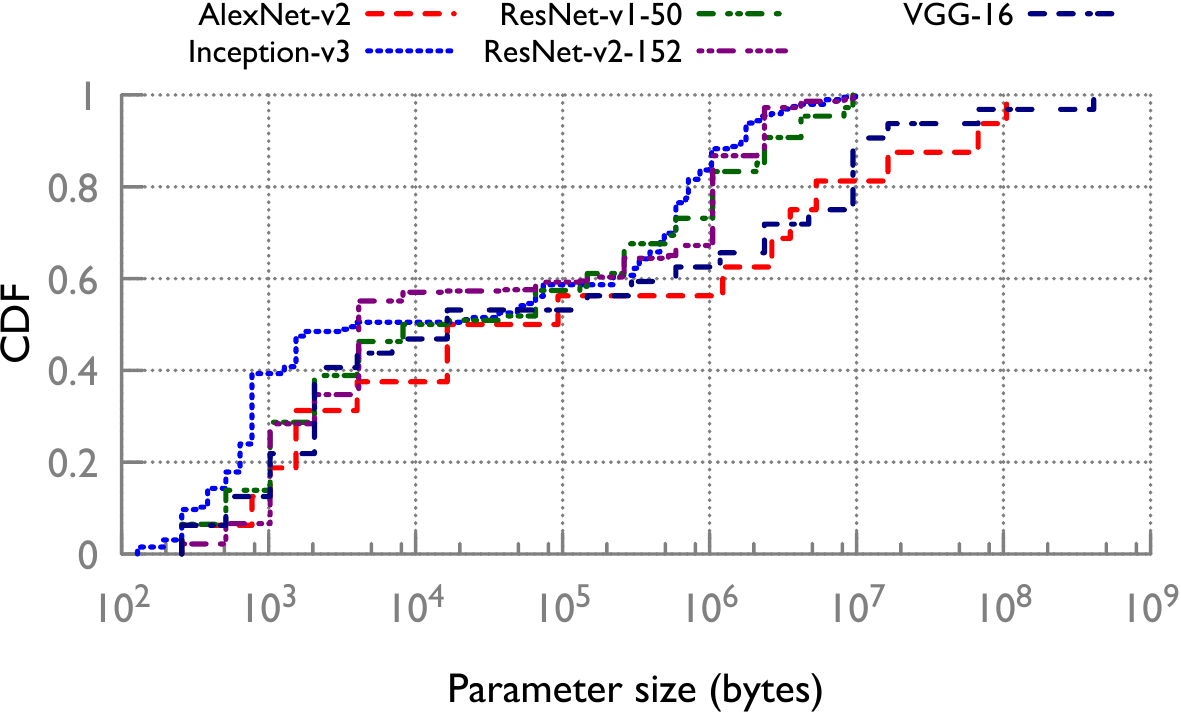}
 \label{fig:parameterCDF}
 \caption{\small Parameter Size distribution in $5$ DNN models}
\end{subfigure}%
\begin{subfigure}[t]{0.28\textwidth}
\centering
\includegraphics[width=\textwidth]{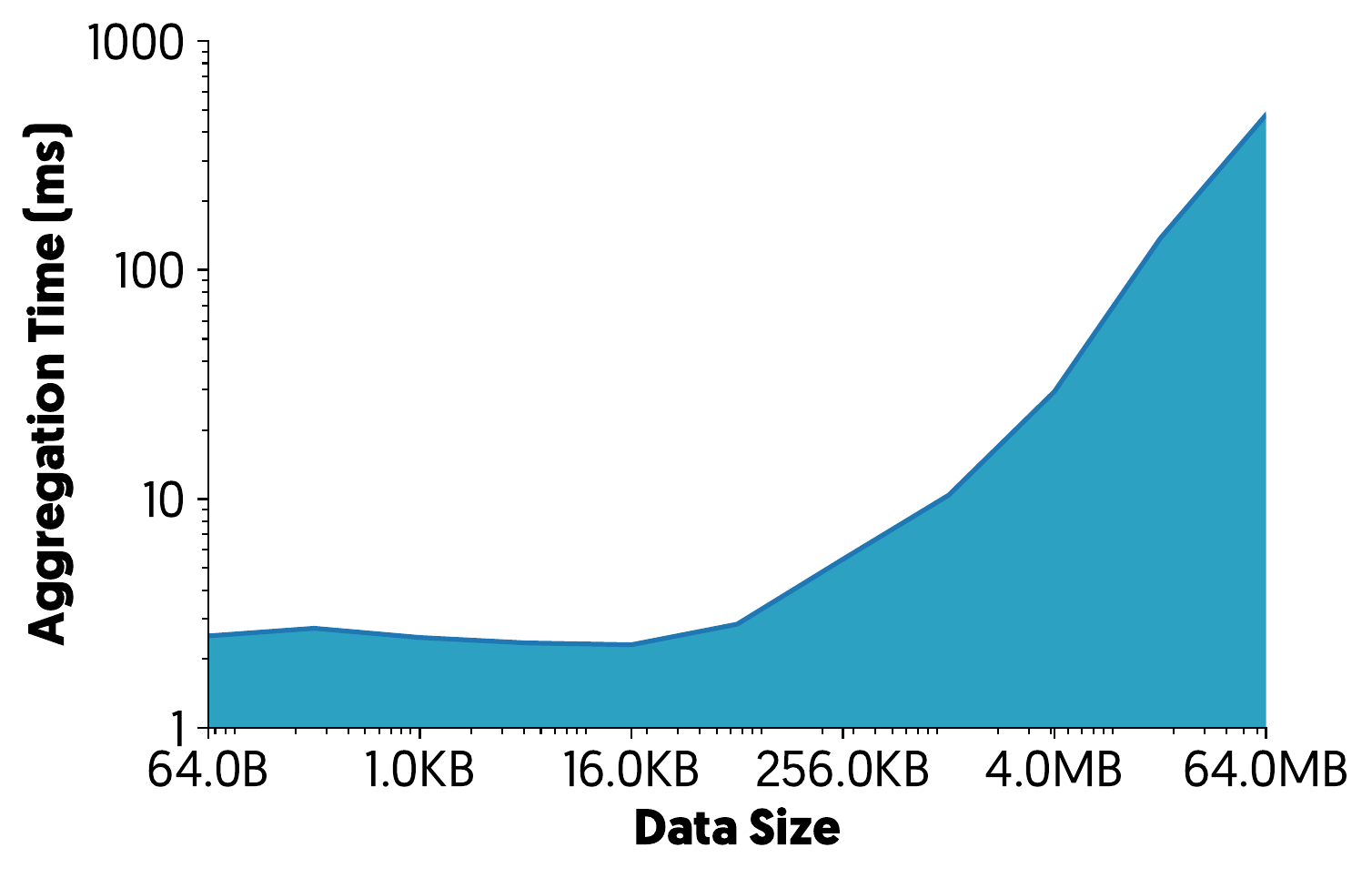}

    \caption{\small End-to-end transfer time within TensorFlow at different data sizes}
\end{subfigure}
\caption{\small Understanding model characteristics: opportunities for optimization (a) significant duration of forward pass which is not used for transfers (b) large number of small parameters, (c) large network overhead at small data sizes}
\label{fig:motive}
\end{figure*}

\subsubsection{Opportunities for Performance Improvement}

\vspace{-1mm}
\parab{(A) Randomness in transfer boundaries:} In decentralized aggregation, all workers should have the parameter available for aggregation before the transfer can be initiated. However, there are multiple feasible orders for executing operations in a DAG. As a result, parameters may become available at different workers in varying orders. For example, Figures~\ref{fig:example_dag_best} and \ref{fig:example_dag_worst} show two schedules of computation operations which are both feasible according to Figure \ref{fig:example_dag}. In the best schedule, transfer boundaries are farther apart, allowing better overlap of computation and communication, which will in turn improve GPU utilization. In the worst schedule, the overlap is significantly reduced due to the shorter window available. Thus, we can increase the window between transfer boundaries through better scheduling of computation operations.

\vspace{-1mm}
\parab{(B) Restrictions on network transfers:} In PS, the parameters are updated at the backpropogation phase of an iteration and read in the forward pass of next iteration. However, current implementations of decentralized schemes restrict these network transfers to the backpropagation phase. As a result, the network is not utilized during the forward pass as shown in Figure~\ref{fig:transfer_window_comparision}. In common models, forward pass accounts for about $30\%$ of the computation time (\autoref{fig:motive}(a)) which is currently not utilized for network transfers.

\vspace{-1mm}
\parab{(C) Large overhead for small parameters:} PDF of parameter sizes across $5$ popular models are given in Figure~\ref{fig:motive} (b). We observe that there are a large number of small parameters, with $50\%$ of parameters smaller than $20$KB in all models. This observation also holds for $15$ other models that we evaluated. Next, we study the impact of small parameters by measuring the time to receive a small parameter within the TensorFlow framework. This is the end-to-end time from the application perspective which includes the network transfer time and the time for serialization/deserialization, kernel to user-space delay etc. In Figure~\ref{fig:motive} (c), we show the end-to-end transfer time from within TensorFlow for different data sizes with recursive doubling-halving algorithm on 16 workers. We observe that small parameters incur a large delay due to non-network overheads. Thus, we can improve performance by batching smaller parameters. Also, different parameters are read and updated at different times, based on their order of activation in the DAG. This opens the door for smart parameter batching and scheduling based on their transfer boundaries.





\begin{figure}
\centering
\begin{subfigure}[t]{0.49\columnwidth}
\centering
  \includegraphics[width=0.9\textwidth]{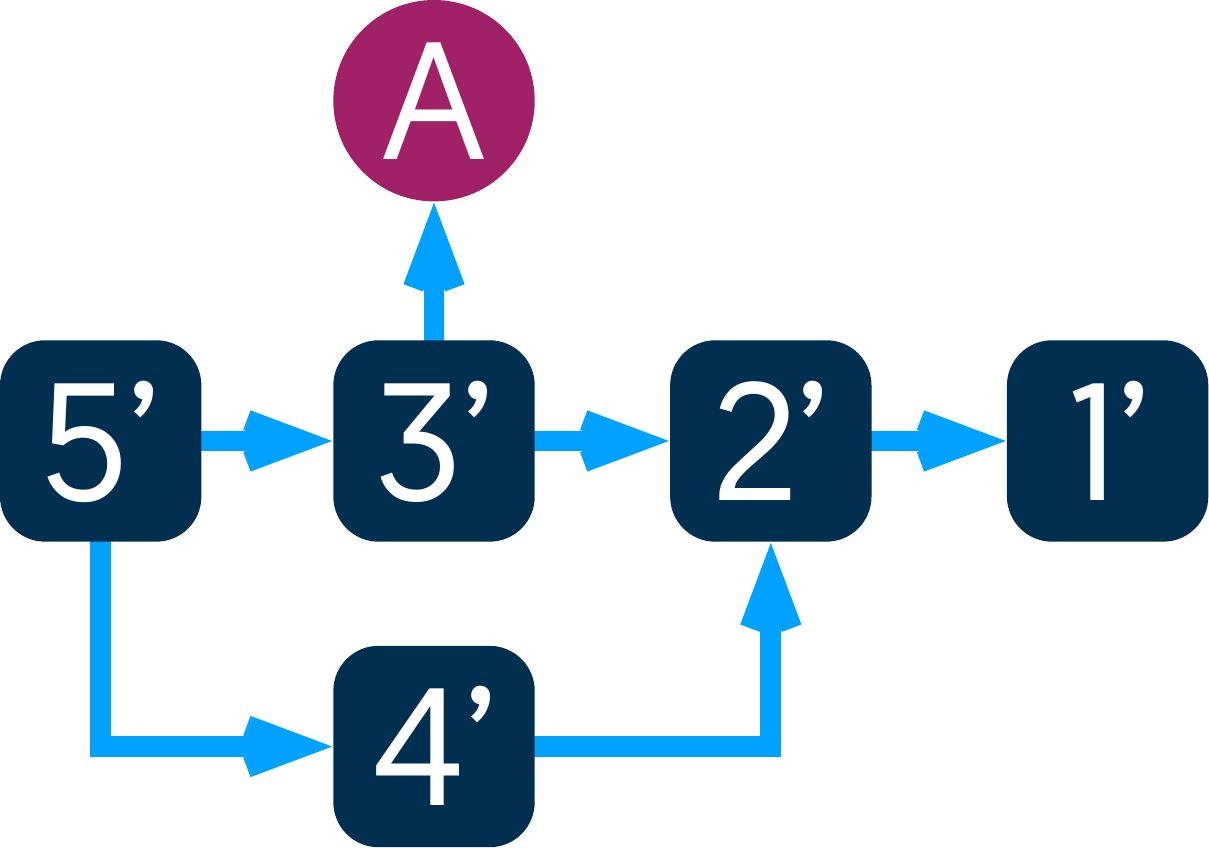}%
  \label{fig:dag_optimization_before}
  \caption{Before}
\end{subfigure}
\begin{subfigure}[t]{0.49\columnwidth}
\centering
  \includegraphics[width=0.9\textwidth]{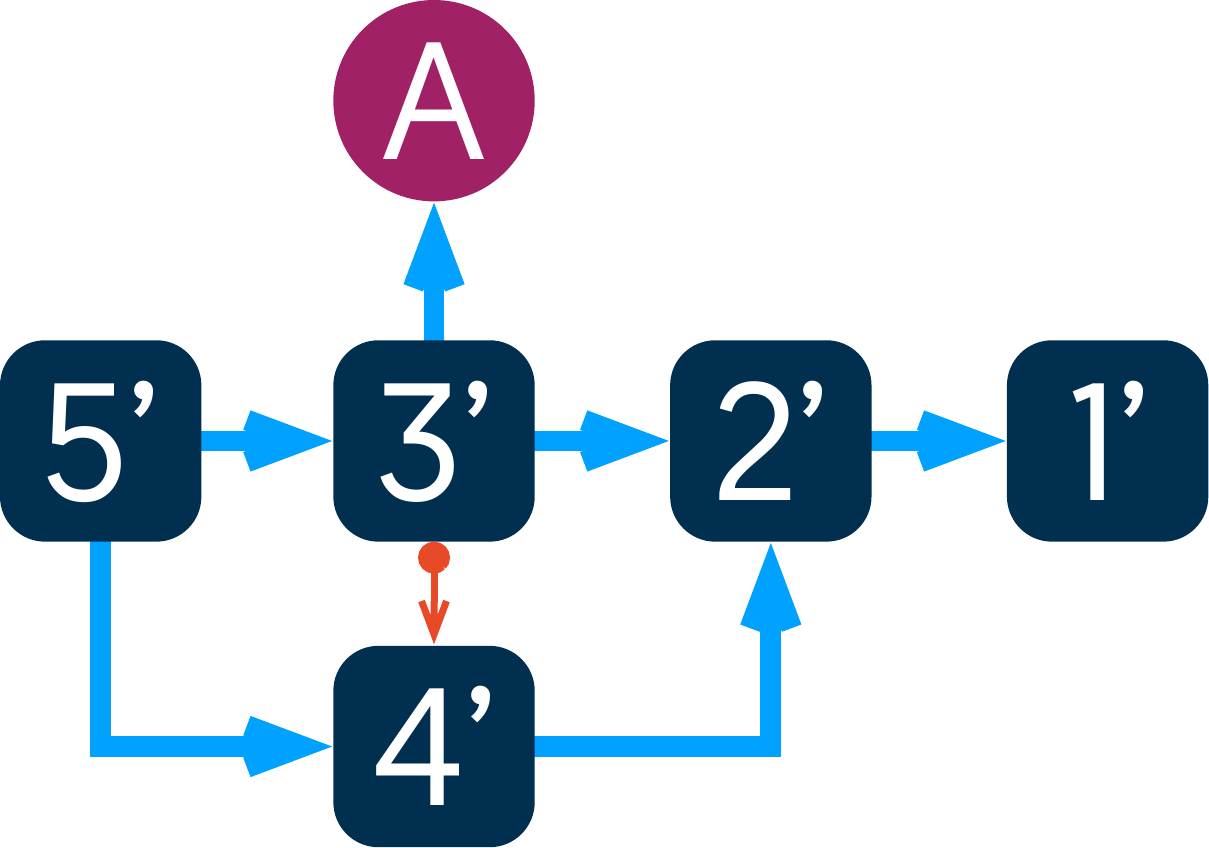}%
  \label{fig:dag_optimization_after}
  \caption{After}
\end{subfigure}
\caption{DAG Optimization in Caramel}
\label{fig:dag_optimization}
\vspace{-3mm}
\end{figure}

\section{\sysname Design}
Network transfer optimization in \sysname involves four functionalities: (i) dataflow DAG optimizer, (ii) small-parameter batcher, (iii) network transfer scheduler, and (iv) adaptive depth enforcer.

\subsection{Dataflow DAG Optimizer}
\label{subsec:dagOptimizer}
In decentralized aggregation patterns, it is necessary to have a parameter ready for aggregation at all workers before it can be aggregated (\S~\ref{sec:modelChar} B). This module is responsible for (i) determining the best executing order of ops in the DAG and (ii) adding additional dependencies in the model to ensure that there is only a single feasible order of execution. 




\begin{figure}
\centering
\begin{subfigure}[h]{\columnwidth}
  \includegraphics[width=0.9\textwidth]{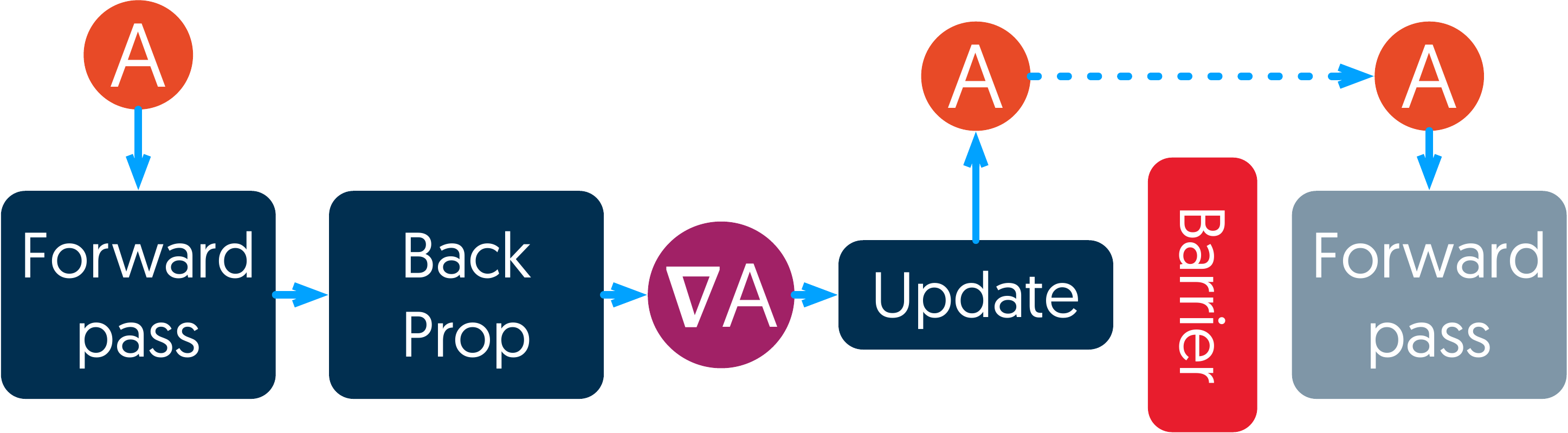}
  \caption{Before}
  \label{fig:scheduling_before}
\vspace{2mm}
\end{subfigure}
\begin{subfigure}[h]{\columnwidth}
  \includegraphics[width=\textwidth]{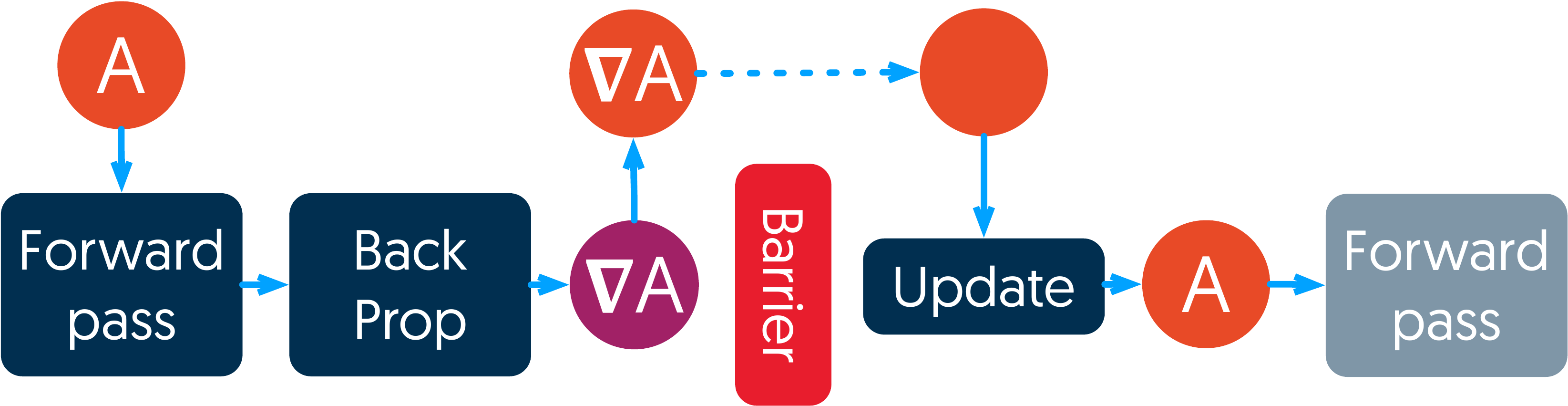}
  \caption{After}
  \label{fig:scheduling_after}
\end{subfigure}
\caption{Transfer Scheduling in Caramel}
\label{fig:scheduling}
\vspace{-2mm}
\end{figure}

\parab{Stage 1 --- Determining the best order: } To maximize the overlap coefficient, $\alpha$, the computations should be prioritized in a manner that activates the communication operations as early as possible (early start boundary for parameters). We add the minimal number of additional dependencies to ensure desired ordering on the parameter updates/activation. 

First, we trace execution of an iteration on a single machine 10 times. The execution time of a computation operation is determined as the minimum observed time across all runs. Empirically, we find that our method can accurately predict the computation time of execution (with less than $3\%$ error in the worst case) with only $5$ runs. 

Next, we use an iterative greedy algorithm to find the best order of parameter updates. In each step, we calculate the total time taken by computation ops that need to be completed before each parameter can be activated. The parameter with the least cost of computation required to activate it is chosen and the computation ops that it depends on are marked as completed (their are not counted as dependencies in the next iterations). This process is repeated until all parameter updates are visited.

\parab{Stage 2 --- Enforcing the best order: }
This is an iterative process where parameters are activated in the best order chosen in the previous stage. In each step, we find the list of all ops that the chosen parameter directly or indirectly depends on. We define the \textit{free set} as the set of all ready-to-execute ops, i.e., ops with no dependencies on any unexecuted ops. The \textit{end set} is the list of ops which the target parameter update depends on directly. New dependencies are added between end set of parameter with tag $i$ and free set of parameter with tag $i+1$. Parameters are executed in the increasing order of their tags (based on the chosen order).


This ensures that at each given time, only ops needed for the target parameter update can be executed. It is worth noting that adding additional control dependencies to the dataflow model does not change the underlying logic of the DAG. The enforced order is one of the feasible orders in which the DAG may be executed, even without the additional dependencies.

\subsection{Parameter Batcher}
\label{subsec:batcher}
Small parameters incur large overhead (\S~\ref{sec:modelChar}). Hence,
the goal of parameter batching is to reduce this overhead by combining parameters in to groups. In our implementation, we focus on grouping small parameters only.  Larger parameters, larger than a certain threshold determined by the network characteristics, are transferred without batching.

We begin with the order we obtained (\S~\ref{subsec:dagOptimizer}) and calculate the expected parameter update time. For each parameter, in the ascending order of the estimated update time, we determine whether to batch it or not. If the size of the parameter is larger than the threshold (the choice of the threshold explained in the next paragraph), it is transferred without batching. If the parameter is smaller than the threshold, we decide whether to transfer immediately (effectively, putting the transfer in a queue to transfer eventually) or add to the current active batch. The current batch is transferred when the active batch size exceeds the threshold or if the transfers in the queue are done before the next parameter update. 

This algorithm ensures that parameters are batched whenever there is an opportunity, i.e., the network is busy with other transfers. The threshold plays an important role in this algorithm. If the threshold is too small, too few parameters will be batched. If it is too large, the batching overhead will exceed the benefit. The threshold can be set manually. 

We use a network model to predict the total transfer time for a given data size. Empirically, we find that a simple linear regression model can accurately predict the transfer time for a given data size in the network. In order to generate this model, we run two sets of network microbenchmarks for two data sizes: $64B$, and $4MB$. The choice of data size is arbitrary; we get very similar results with different combinations. For each chosen data size, we run sequential aggregation transfers and record the time. Next, we fit this data to a linear model. Using theis network model, we estimate the best threshold as follows:
$$
Threshold = min_{x} \frac{f(2x)}{2f(x)} > 0.8
$$
where $f(d)$ is the network transfer time for data size $d$. We obtain the minimum overhead using this threshold. 

\subsection{Network Transfer Scheduler}
The network transfer scheduler is responsible for increasing the overlap coefficient by scheduling parameter transfers efficiently. Transfers are scheduled in both backward pass and forward pass to overcome the shortcomings discussed in \S~\ref{sec:modelChar}, without affecting the computation (Figure~\ref{fig:scheduling}).

Moving a network transfer to the forward pass has the possibility of causing delay in computation. We avoid this problem through model-awareness. A parameter cannot be updated beyond its transfer boundaries. For batched parameters, this boundary is determined by the parameter that is read at the earliest time/updated at the latest time. 


We implement a greedy $2$D-Bin packing algorithm to pack the parameters based on their feasible window. The two dimensions are time and data size. The algorithm proceeds as follows. First, we sort the batched groups in the descending order of size. Second, for each group, we attempt to pack the group in parallel with computation, first in the backward pass (any time after the end of the batch); if that is not feasible, next, in the forward pass (any time before the start of the batch). At the end of this stage, we have a few groups which are allocated a transfer time and some that are unassigned.

In the third stage, we repeat the same process on unassigned items, but allowing transfers beyond the computation time, i.e., after the end of the backward pass, or before the start of the forward pass. For each item, we compare the additional time added to iteration time by placing the group in FP and BP, and we choose the one with the smallest overhead.

\subsection{Adaptive Depth Enforcer}
In decentralized algorithms, there are two or more stages where data is transferred and aggregated across participating nodes. In each step, data is transferred on the network, and is sent to application to be reduced, before the result is sent again over the network. This back and forth between network and application reduces the network utilization since the network is not utilized during the reduction at the application layer. One solution for avoiding this  network under-utilization is to chunk (or break) the data in to a few pieces, and transfer each chunk independently in parallel. The number of chunks is called \textit{depth} of algorithm. In this case, while one chunk is being reduced on the CPU, another chunk can be sent over the network, i.e., this enables pipelining of network transfer and application-level processing across various chunks. The choice of depth in some DL systems is fixed. For example in \cite{goyal2017accurate} a fixed depth of 2 is used. Throughout the experimentation we observe that the depth has a conflicting effect on transfer performance. As shown in Figure~\ref{fig:depth}, transfer time of small parameters increases with increasing depth. In the worst case, we observe $3\times$ slow down going from depth of $1$ to $8$. For large parameters, however, the transfer time decreases by increasing the depth. At the peak, we observer $60\%$ decrease in transfer time going from depth of $1$ to $8$. 

We choose the depth of transfer adaptively, starting from a depth of $1$ at smaller parameter sizes to a maximum of $8$ at larger sizes. The depth is determined based on the data size and a threshold, (this is same as the parameter batching threshold in~\ref{subsec:batcher}). As shown in Figure~\ref{fig:depth}, our adaptive depth gets the best performance; smallest transfer time at all sizes.

\vspace{-2mm}
\paragraph{In-Graph Implementation}
In contrast to other implementations of decentralized aggregation in deep learning systems such as Horovod (in TensorFlow and PyTorch) and Gloo (in Caffe2) where the aggregation pattern is abstracted as a single op in the dataflow model, we implement the aggregation pattern as a part of the DAG. In other words, the data conversion, transfer, and aggregation are defined as standard dataflow ops. This allows the aggregation pattern to take advantage of further optimizations by the framework such as op fusing, XLA \cite{leary2017xla}. Additionally, in-Graph implementation does not dependent on external dependencies such as MPI, making it more accessible and easier to deploy in the cloud environment. 

\section{Implementation}
We implement \sysname as a Python library over TensorFlow. The code is publicly available (obfuscated for review). The library takes user code dataflow model intended for a single device, and generates an In-Graph distributed dataflow model. \sysname API declaration is as follows:
\vspace{-2mm}
\begin{verbatim}
def ARModel(context,
            number_of_workers,
            serialization = True,
            batching = True,
            scheduling = True,
            analyzers = None,
            device_list = None)
\end{verbatim}

The functionality of \sysname is divided in two: 1) extracting information from the environment and calculating the best network schedule based on the user-provided code, 2) generating a new distributed TensorFlow dataflow model with the added optimizations.

\begin{figure}[t]
  \centering
  \includegraphics[width=0.9\columnwidth]{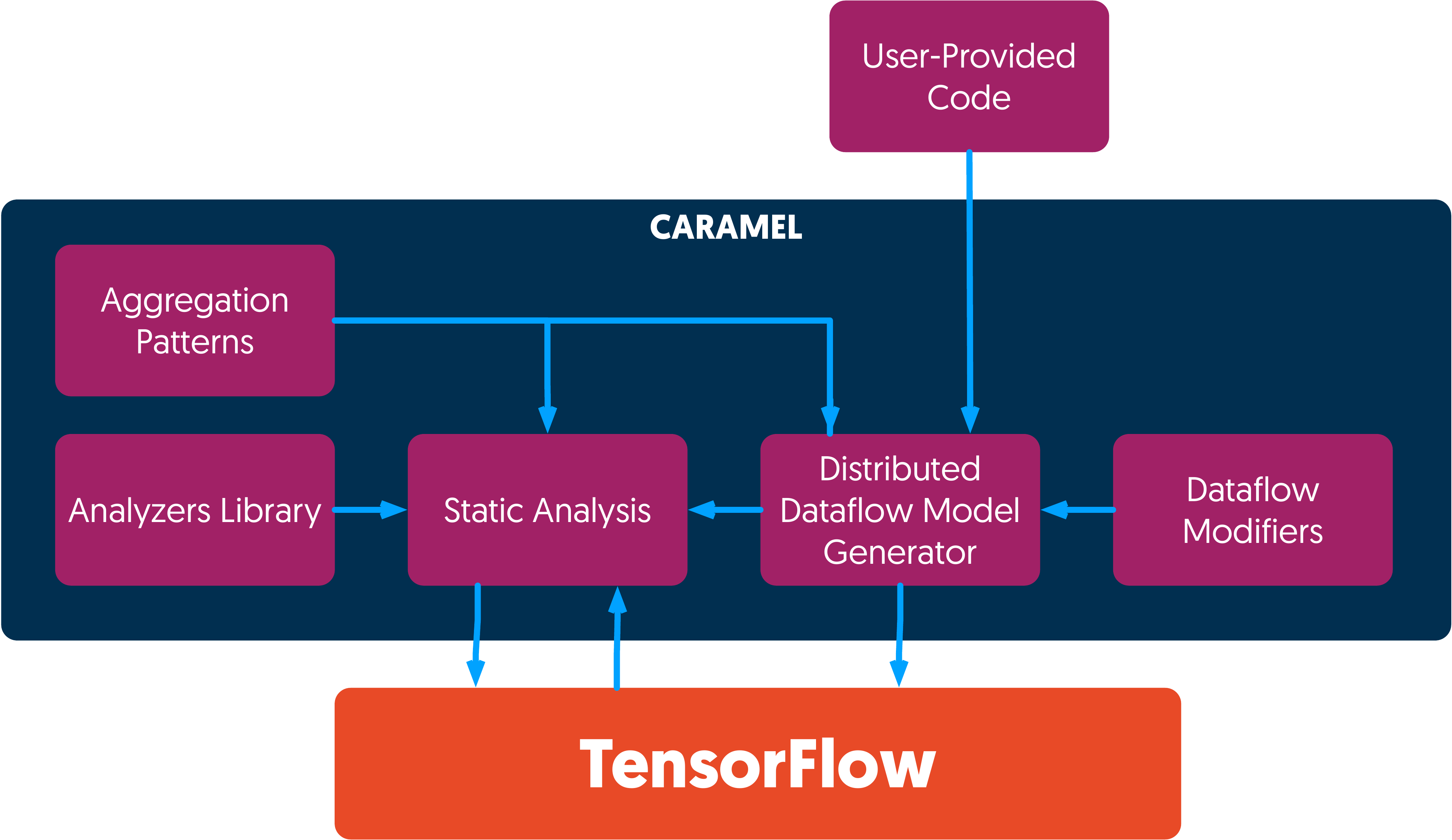}
  \caption{Caramel Implementation}
  \label{fig:system-arch}
  \vspace{-2mm}
\end{figure}

Figure~\ref{fig:system-arch} shows the main components of Caramel. 
\textbf{Distributed Dataflow Model Generator} is the component which glues together all the other components in the system and provides an interface to the user to interact with \sysname. The ultimate goal of this component is to generate a network-optimized distributed model. This component generates a distributed dataflow model using the \textbf{aggregation pattern} as the network primitive. Next, it applies the optimizations on the dataflow model through \textbf{Dataflow Modifiers}. Each modifier applies an optimization on the dataflow graph. For example, the DAG optimizer takes a list of control dependencies and adds it to the dataflow model. The behavior of the modifiers and the choice of aggregation pattern is controlled by the \textbf{analyzers}. Each analyzer generates a piece of information to be used by other analyzers or modifiers. \textbf{Static Analysis} component is responsible for figuring out the data dependencies between analyzers and executing them. The \textbf{Distributed Dataflow Model Generator} automatically selects the set of analyzers based on optimizations. However, the user can send a custom list of analyzers.



\section{Experiments}
In this section, we evaluate the efficiency of Caramel system implemented over TensorFlow. 


\begin{figure*}[h]
\centering
\begin{subfigure}[t]{0.45\textwidth}
\centering
  \includegraphics[width=0.9\textwidth]{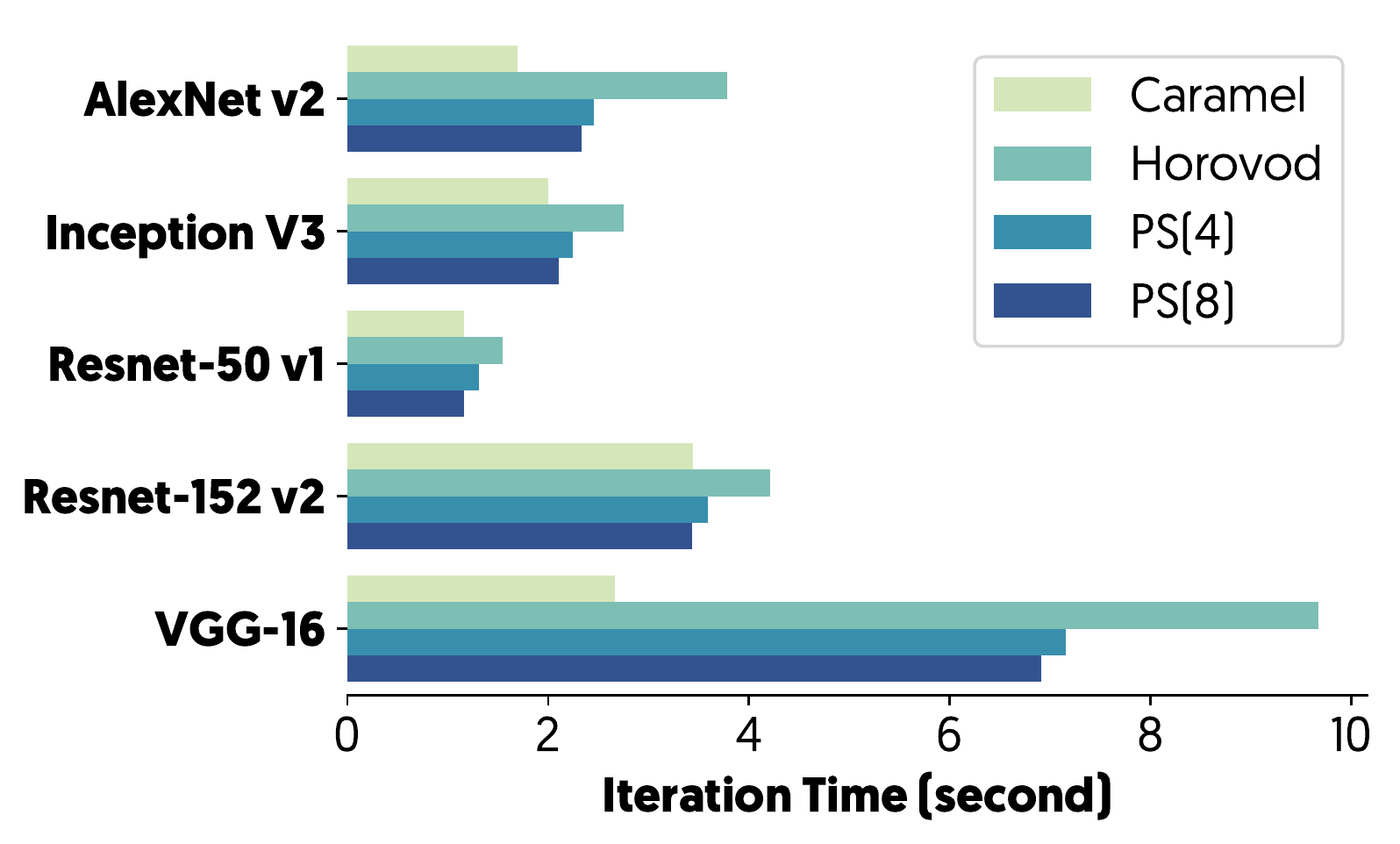}
  \label{fig:iteration-time-8}
  \caption{8-Worker}
\end{subfigure}
\begin{subfigure}[t]{0.45\textwidth}
\centering
  \includegraphics[width=0.9\textwidth]{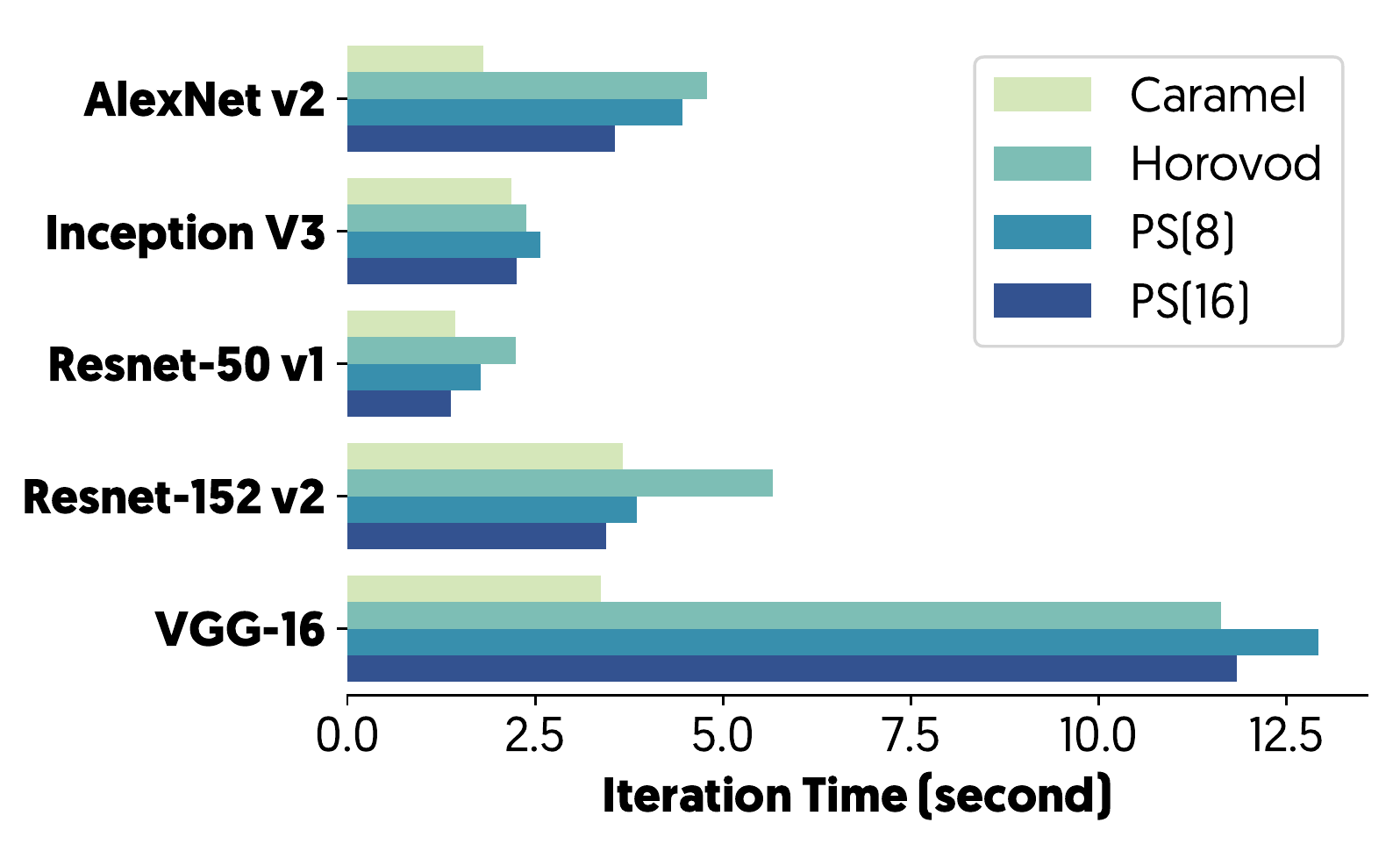}
  \label{fig:iteration-time-16}
  \caption{16-Worker}
\end{subfigure}
\caption{Comparison of Iteration Time in Caramel with PS and Horovod. Lower is better.}
\label{fig:iteration-time}
\end{figure*}



\begin{figure*}
\centering
\begin{subfigure}[t]{0.45\textwidth}
\centering
  \includegraphics[width=\textwidth]{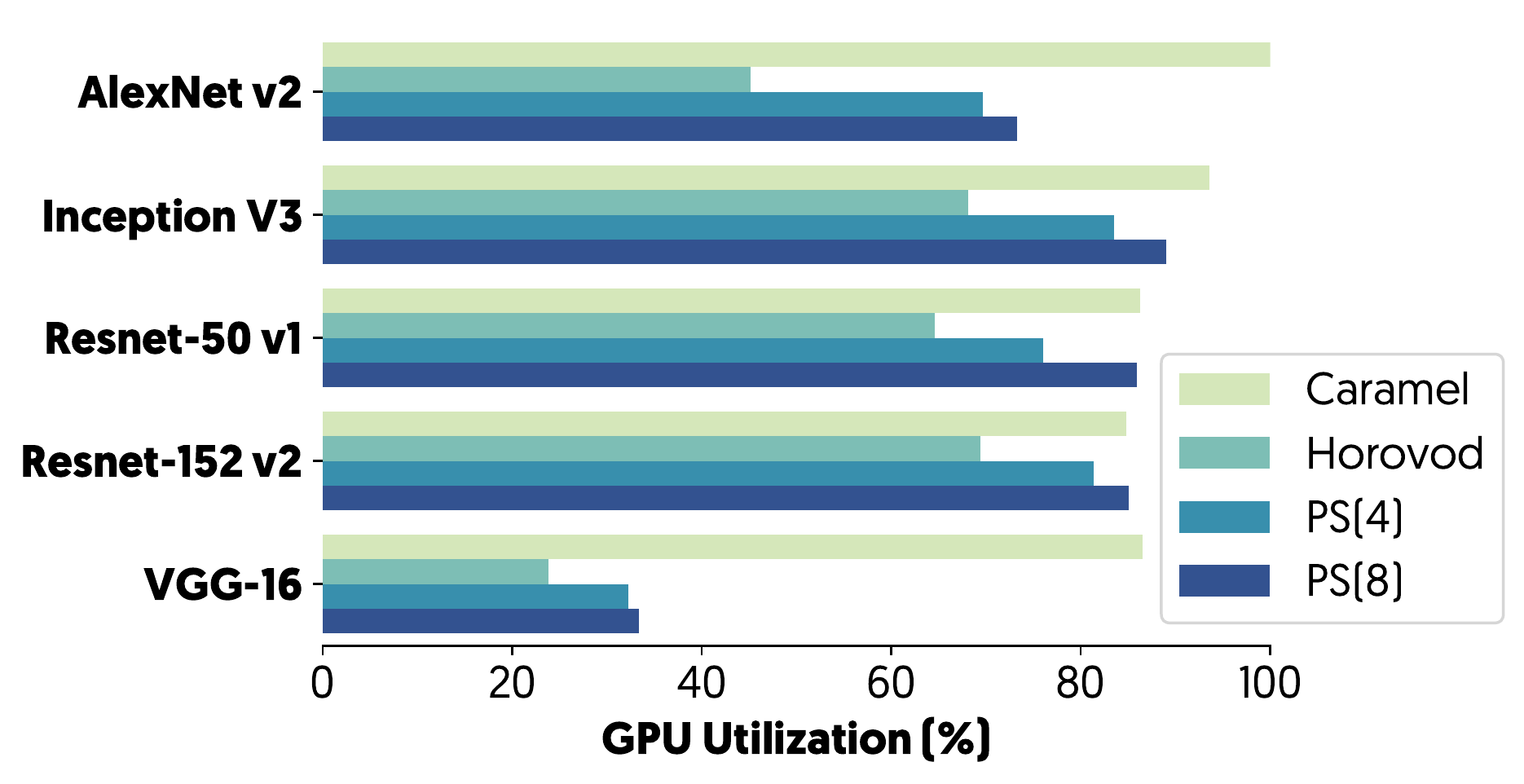}
  \label{fig:gpu-util-8}
  \caption{8-Worker}
\end{subfigure}
\begin{subfigure}[t]{0.45\textwidth}
\centering
  \includegraphics[width=\textwidth]{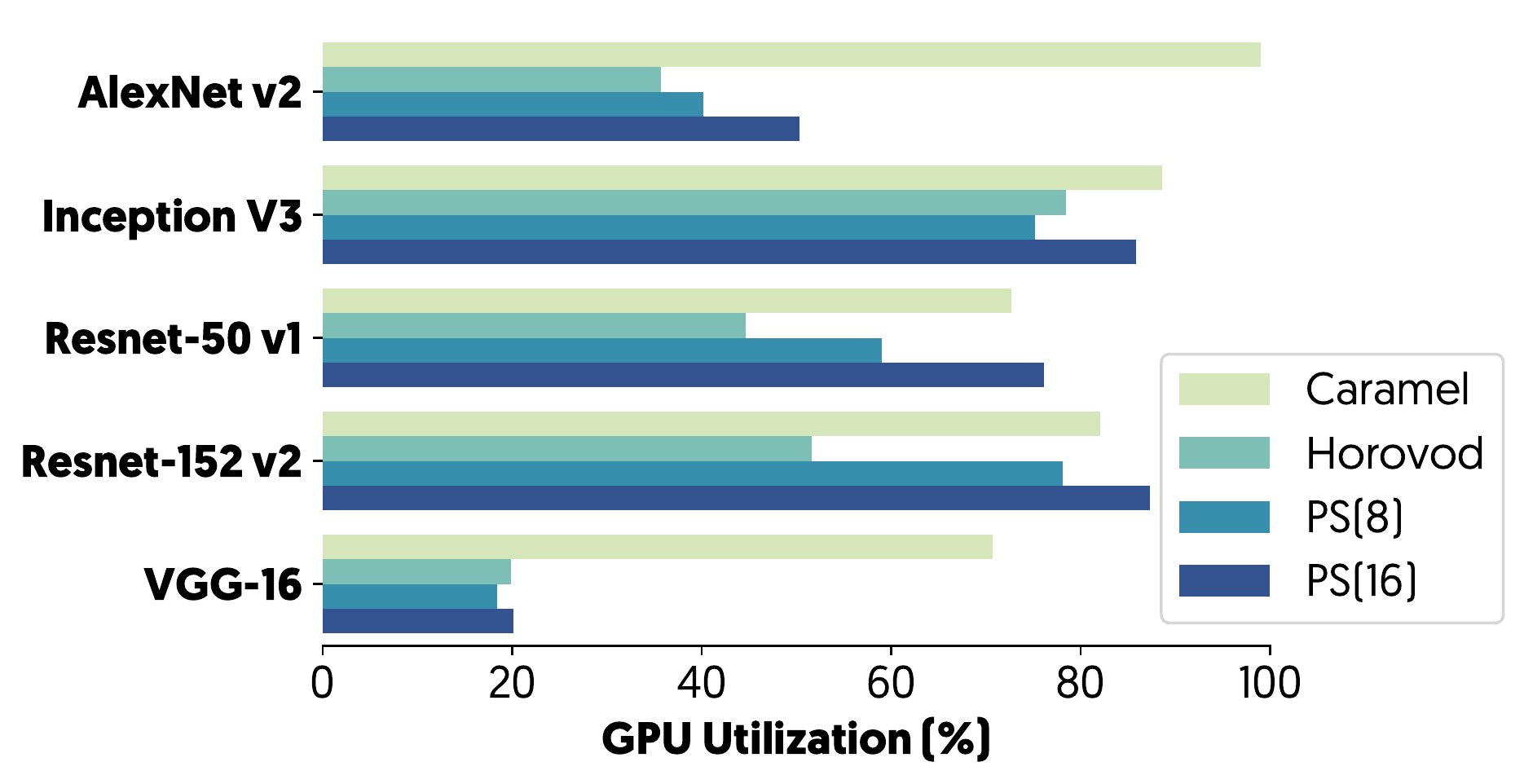}
  \label{fig:gpu-util-16}
  \caption{16-Worker}
\end{subfigure}
\caption{Comparison of GPU Utilization in Caramel with PS and Horovod. Higher is better.}
\label{fig:gpu-util}
\end{figure*}

\begin{figure*}
\centering
\begin{subfigure}[t]{0.45\textwidth}
  \centering
  \includegraphics[width=0.9\textwidth]{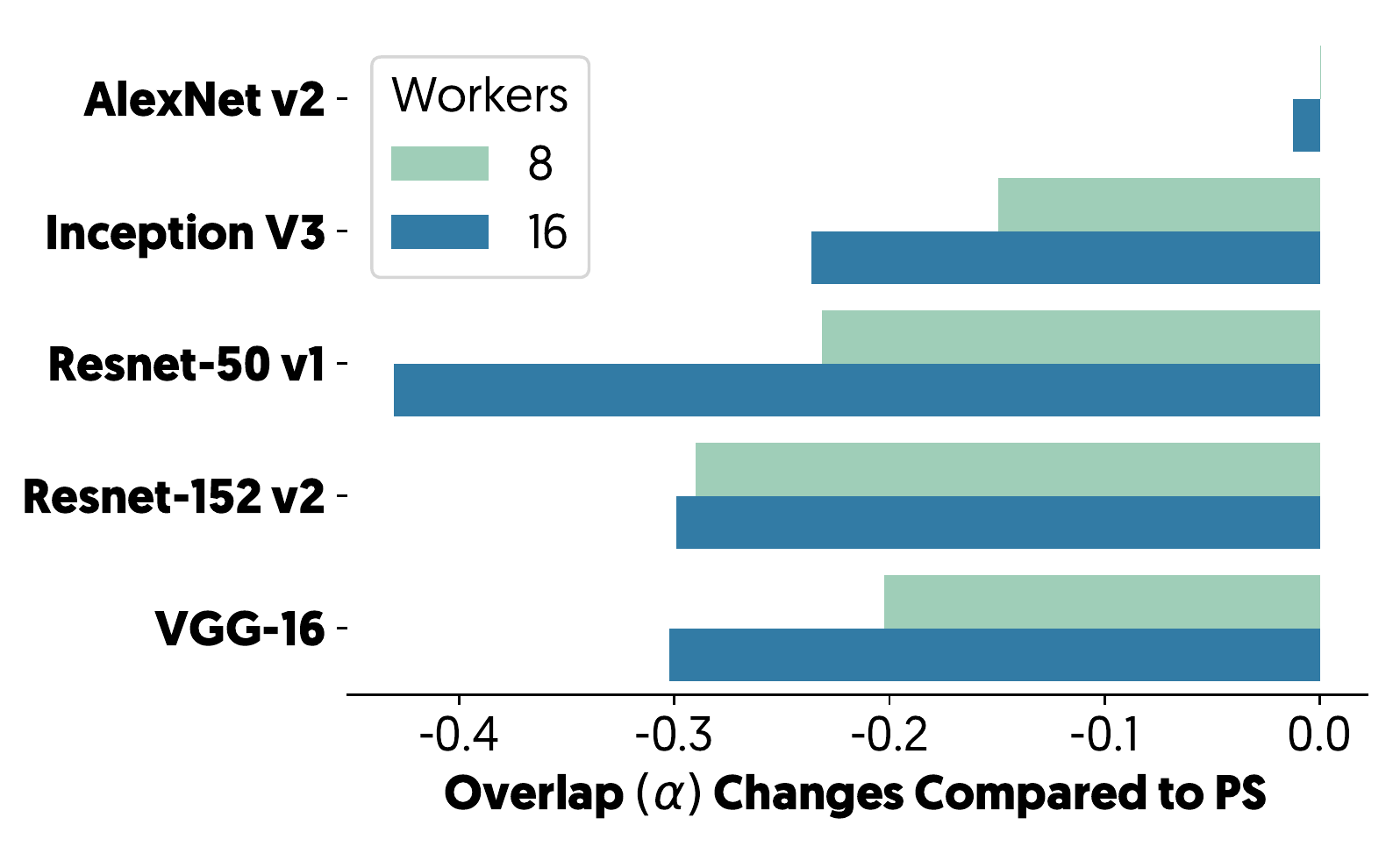}
  \label{fig:vs_ps_overlap}
  \caption{Overlap vs PS}
\end{subfigure}
\begin{subfigure}[t]{0.45\textwidth}
  \centering
  \includegraphics[width=0.9\textwidth]{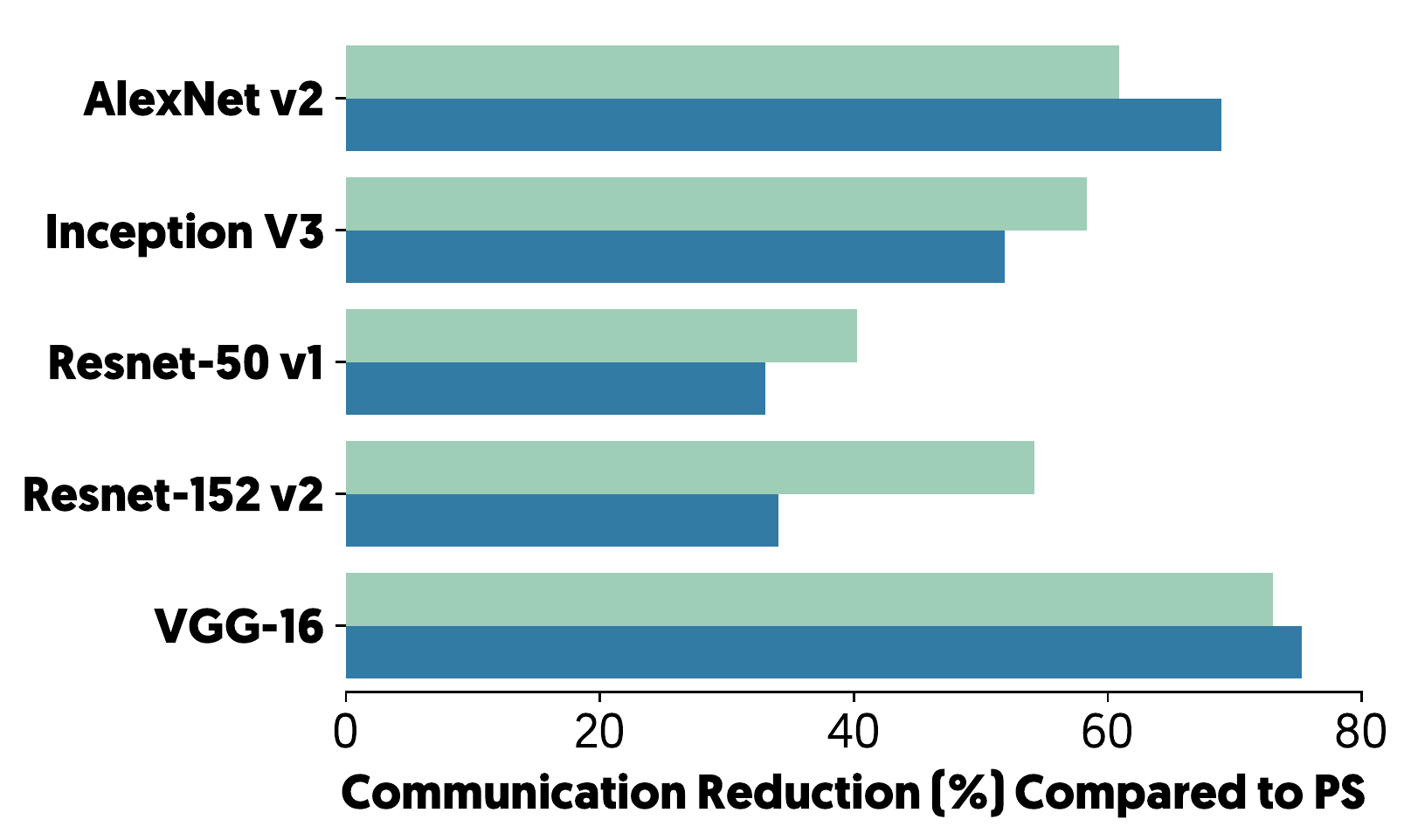}
  \label{fig:vs_ps_net}
  \caption{Network Cost vs PS}
\end{subfigure}
%
%
\begin{subfigure}[t]{0.45\textwidth}
  \centering
  \includegraphics[width=0.9\textwidth]{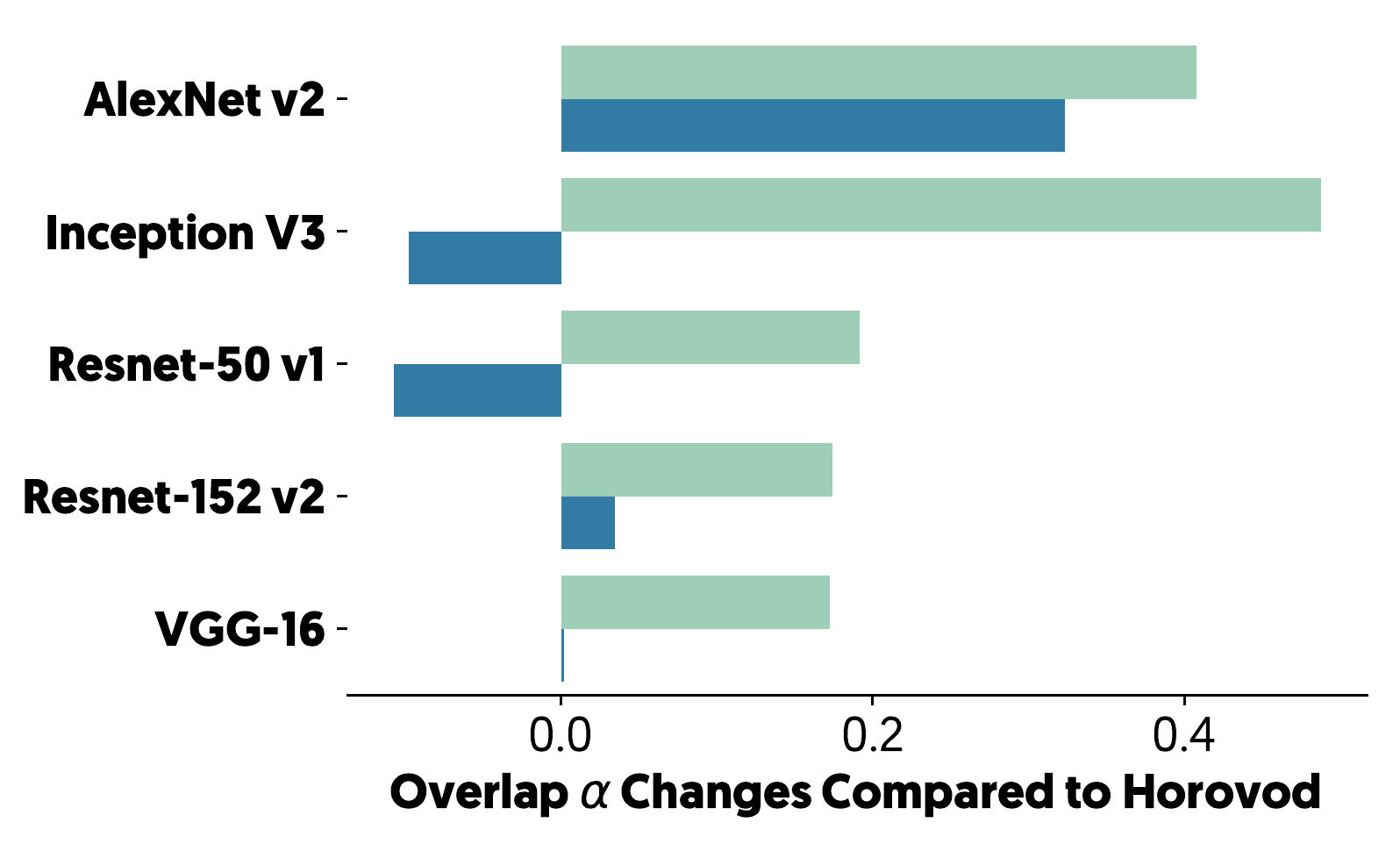}
  \label{fig:vs_ar_overlap}
  \caption{Overlap vs Horovod}
\end{subfigure}
\begin{subfigure}[t]{0.45\textwidth}
  \centering
  \includegraphics[width=0.9\textwidth]{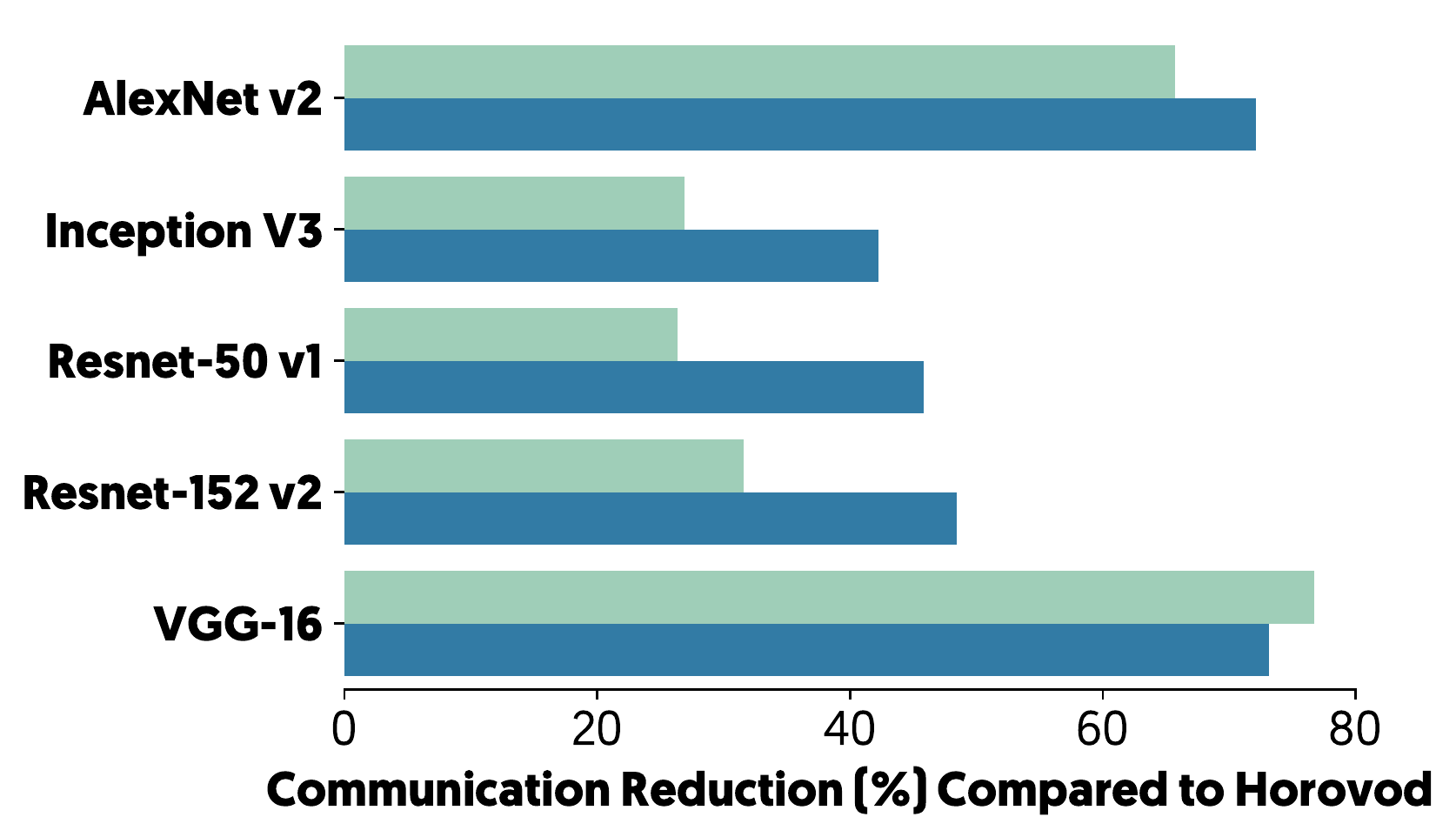}
  \label{fig:vs_ar_net}
  \caption{Network Cost vs Horovod}
\end{subfigure}
\caption{Micro Comparison of Caramel with PS and Horovod. Higher is Better.}
\label{fig:vs_ar}
\vspace{-2mm}
\end{figure*}

\parab{Experiment settings: } We run our tests on Azure cloud environment using Standard NC6 virtual machines (6 cores, 56 GB memory, 1 X Nvidia K80 GPU with 12GB memory). The bandwidth is $10$ Gbps. Our evaluations use $8$ to $16$ workers.

We use \textit{Microsoft Data Science Virtual Machine for Linux (Ubuntu)} image on our VMs which comes with CUDA 9.0, cuDNN 7.0.5, Anaconda Python 3.5.4, Open MPI 1.10.2, and Horovod 0.11.3. We upgrade the TensorFlow to the GPU-enabled binary release of 1.8 from $pip$ repository. 

\parab{DNN models: } We analyze $16$ models and select $5$ representative neural networks for our experiments. (Model, number of parameters, total parameter size (MiB)) are as follows: (AlexNet-v2~\cite{krizhevsky2014one}, 16, 191.9), (Inception-v3~\cite{DBLP:journals/corr/SzegedyVISW15}, 196, 103.5), ResNet-v1-50~\cite{DBLP:journals/corr/HeZRS15}, 108, 97.4), (ResNet-v2-152~\cite{DBLP:journals/corr/HeZR016}, 363, 229.5), and (VGG-16~\cite{simonyan2014very}, 32, 527.8). We use the reference implementation in \url{github.com/tensorflow/models}. 


We evaluated both synthetic and real data based training. For real data, we read the Imagenet Dataset in TFRecord format from a shared NFS-connected Azure storage, resize it with augmentation and prefetch the data during the training. This initial evaluation showed that we have less than $1\%$ iteration time difference between experiments with synthetic data and real data (except in AlexNet-v2 with $3\%$ error). Hence, for the rest of the experiments, we rely on synthetic data.

\subsection{Comparison with other systems}
We compare performance of \sysname with Parameter Server scheme (with \#Parameter servers = \#workers and \#workers/2) and Horovod (state-of-the-art decentralized aggregation scheme). We evaluate two metrics: iteration time (Figure~\ref{fig:iteration-time}) and GPU utilization (Figure~\ref{fig:gpu-util}) with $8$ and $16$ workers.  We observe that performance of \sysname is consistently better than PS and Horovod with lower iteration time and higher GPU utilization across all configurations tested. The largest improvement is observed with VGG-$16$ at $16$ workers with $3.62\times$ improvement in iteration time and $3.5\times$ in GPU utilization. This highest benefit is observed for DNNs with largest variance in parameter sizes. We also observe that \sysname optimizations result in a GPU utilization of atleast $70\%$ in all networks tested.

To understand the performance better, we trace the execution of each iteration using \texttt{tensorflow-tracer}~\cite{tensorflow-tracer}. We measure $\alpha$ and $\rho$ from the traces (Inception-v3 example in \autoref{fig:inception_example}). In Figure~\ref{fig:vs_ar}, we observe that at all sizes tested, \sysname results in reduced communication cost compared to the baselines due to adaptive depth and batching. The benefits accrued by \sysname over PS is due to reduced network cost and over Horovod is due to better overlap. While overlap of \sysname is better than Horovod, it is still worse than PS. However, this is compensated by significant reduction in network cost.

\subsection{Impact of Caramel Optimizations}
\begin{figure}
  \centering
  \includegraphics[width=\columnwidth]{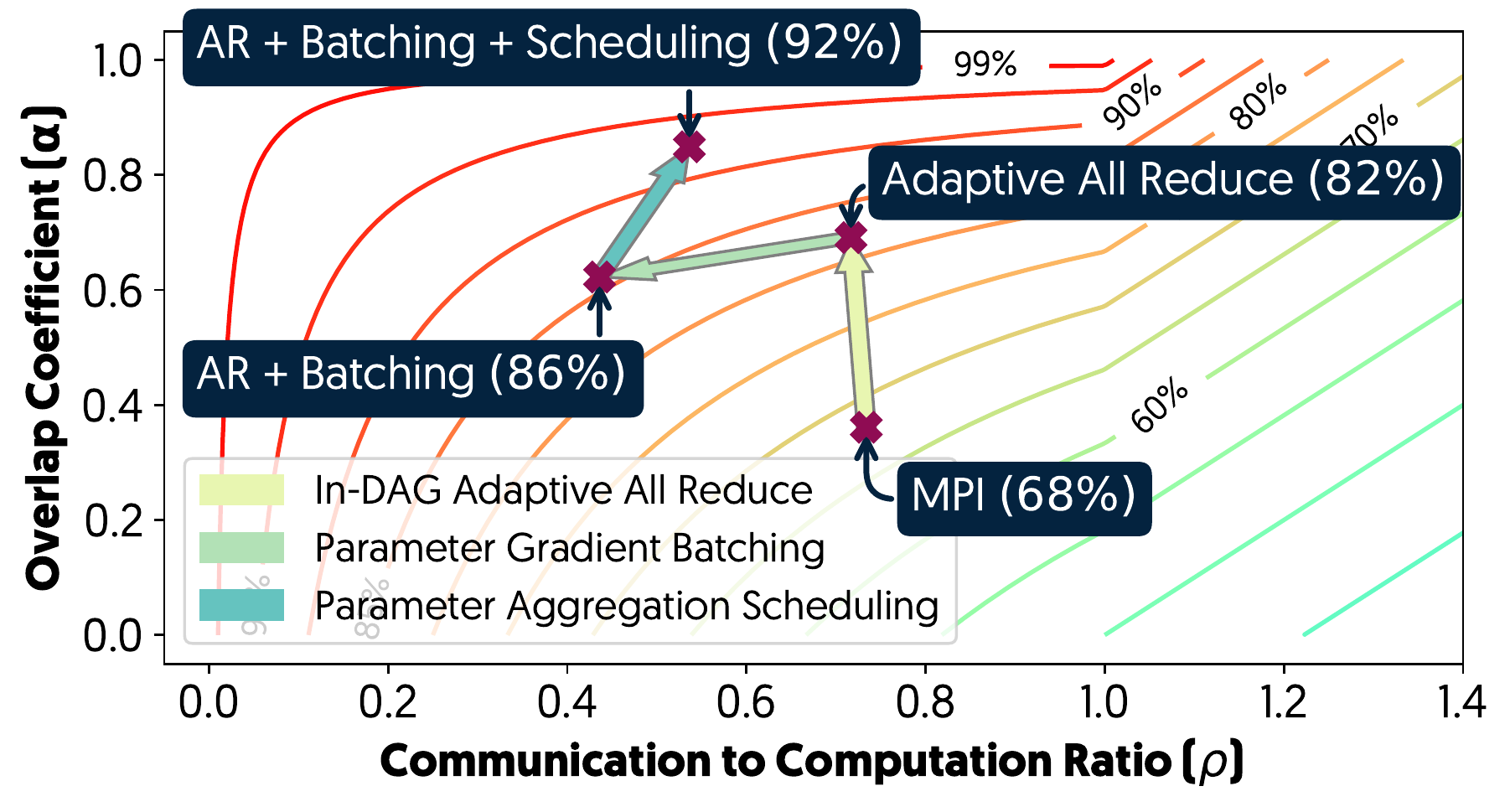}%
  \caption{Caramel Components Contributions on Performance of Inception v3}
  \label{fig:components_inception}
  \vspace{-2mm}
\end{figure}

\begin{figure*}[h]
\centering
\begin{subfigure}[t]{0.45\textwidth}
  \centering
  \includegraphics[width=0.9\textwidth]{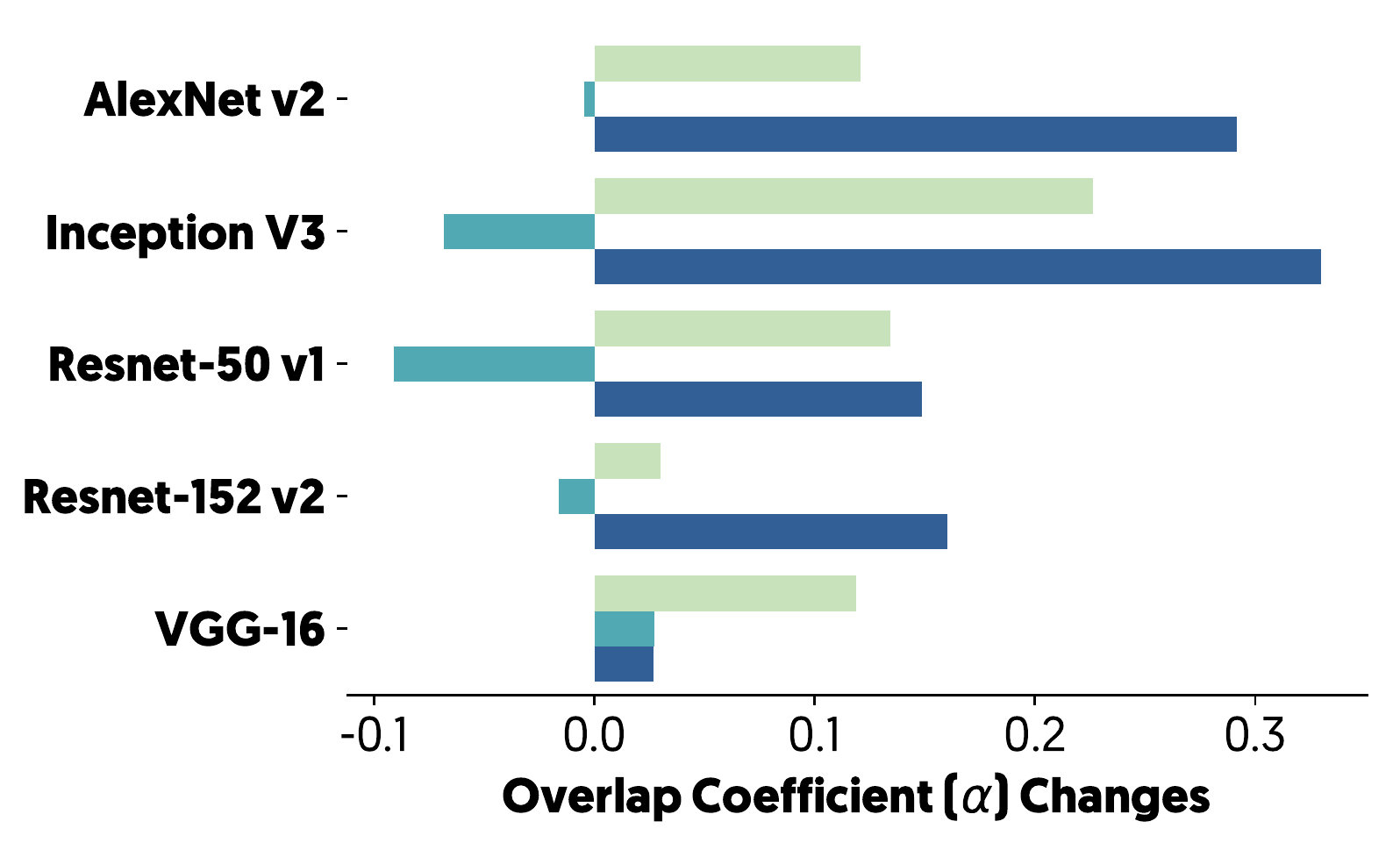}
  \label{fig:components_overlap}
  \caption{Overlap}
\end{subfigure}
\begin{subfigure}[t]{0.45\textwidth}
  \centering
  \includegraphics[width=0.9\textwidth]{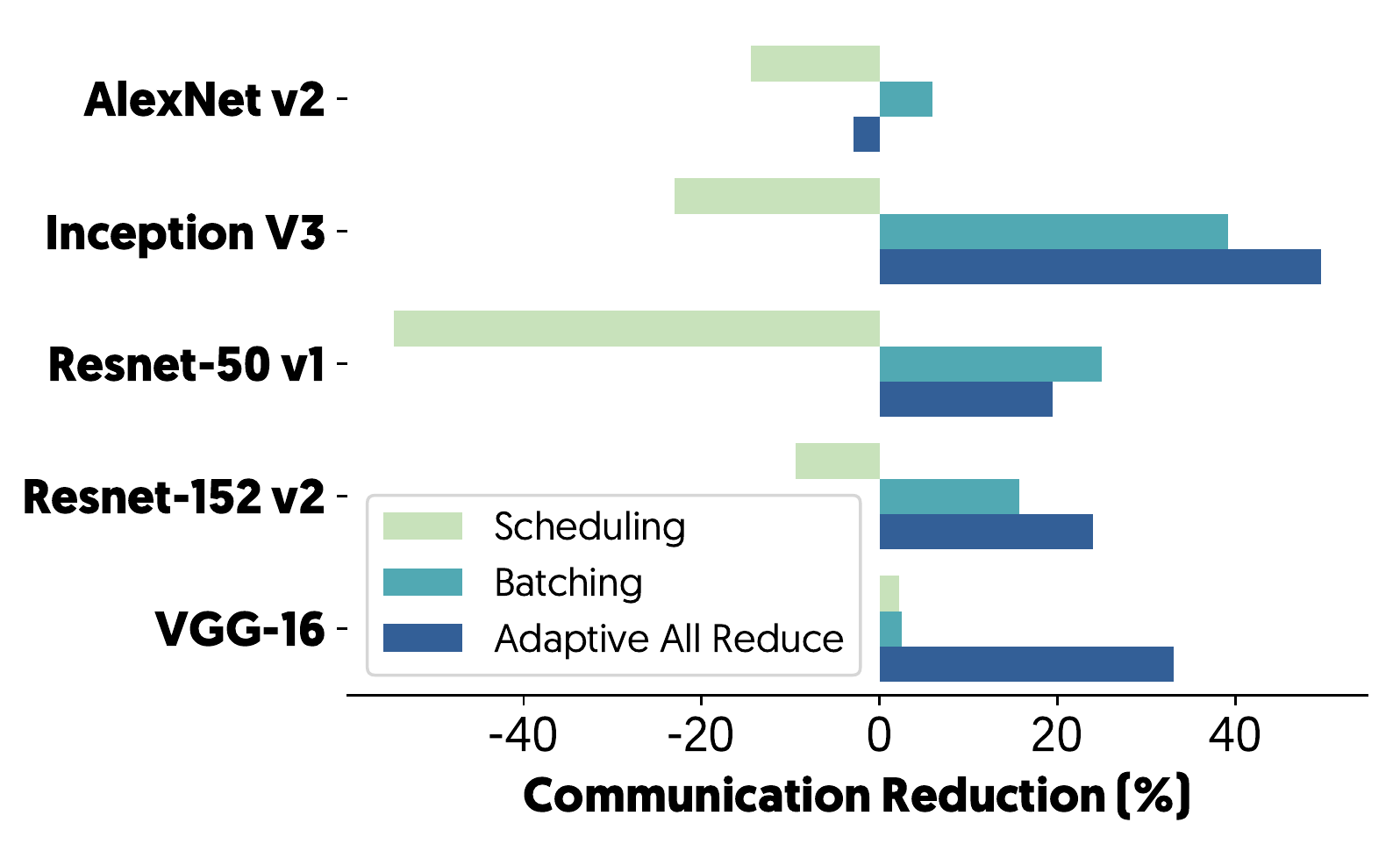}
  \label{fig:components_net}
  \caption{Network Cost}
\end{subfigure}
\caption{Contribution of each module in Caramel to performance. Higher is better.}
\label{fig:components}
\end{figure*}

In this section, we quantify the contribution of each of the optimizations in terms of overlap coefficient ($\alpha$) and communication cost in \sysname towards the performance benefits achieved. In Figure~\ref{fig:components_inception}, we see the impact of putting these optimizations together on a single model, Inception-v3 with $8$ workers. \textit{\textbf{Adaptive All Reduce:}} Compared to the MPI implementation in Horovod, the adaptive all reduce significantly improves the overlap and cost, thereby the GPU utilization by $14\%$.
\textit{\textbf{Batching:}} Implementation reduces the overlap slightly. However, significant reduction in communication overhead further improves the GPU utilization by $4\%$. \textit{\textbf{Transfer Boundaries:}} Adding the computation scheduling, increases communication cost, but improves overlap significantly. As a result, the GPU utilization increases by $24\%$ ($92\%$ in \sysname vs. $68\%$ in MPI implementation of Horovod). 
Similar trends hold in training of other networks as shown in Figure ~\ref{fig:components}.


\begin{figure}
    \centering
    \includegraphics[width=0.9\columnwidth]{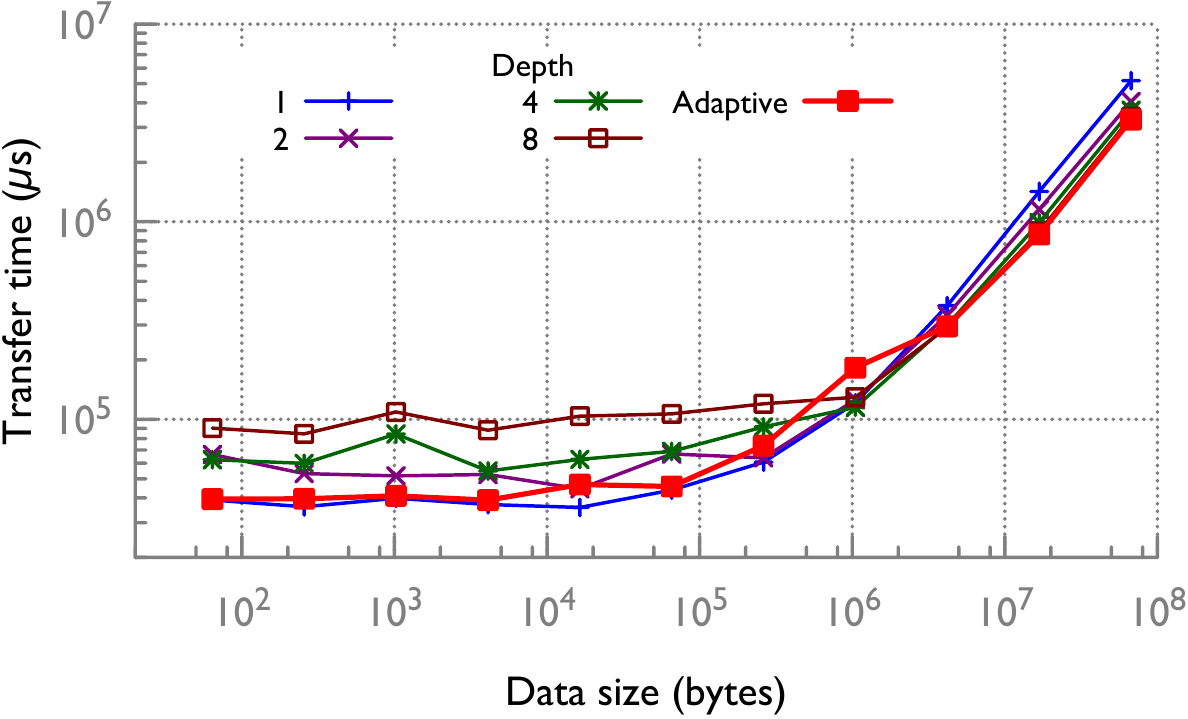}
    \caption{Impact of depth on transfer time}
    \label{fig:depth}
    \vspace{-2mm}
\end{figure}

\subsection{Evaluation of Adaptive Decentralized Schemes}


We test constant depth and adaptive depth schemes at different data sizes (shown in Figure~\ref{fig:depth}. At smaller data sizes, splitting the data to be aggregated into smaller chunks results in increased transfer time. This is caused by the high overhead in each chunk. Hence, a smaller value of depth works better for small data transfers. At large data sizes, on the other hand, a larger depth allows pipelining of multiple transfers, particularly the processing at nodes. The adaptive scheme in \sysname chooses depth of $1$ at small data sizes and a depth of $8$ at the largest size tested. Note that the y-axis is logscale; the adaptive scheme achieves $~60\%$ lower transfer time compared with depth $1$ at $100$ MB.

\begin{figure}
\centering
\begin{subfigure}[t]{0.42\textwidth}
  \centering
  \includegraphics[width=0.8\textwidth]{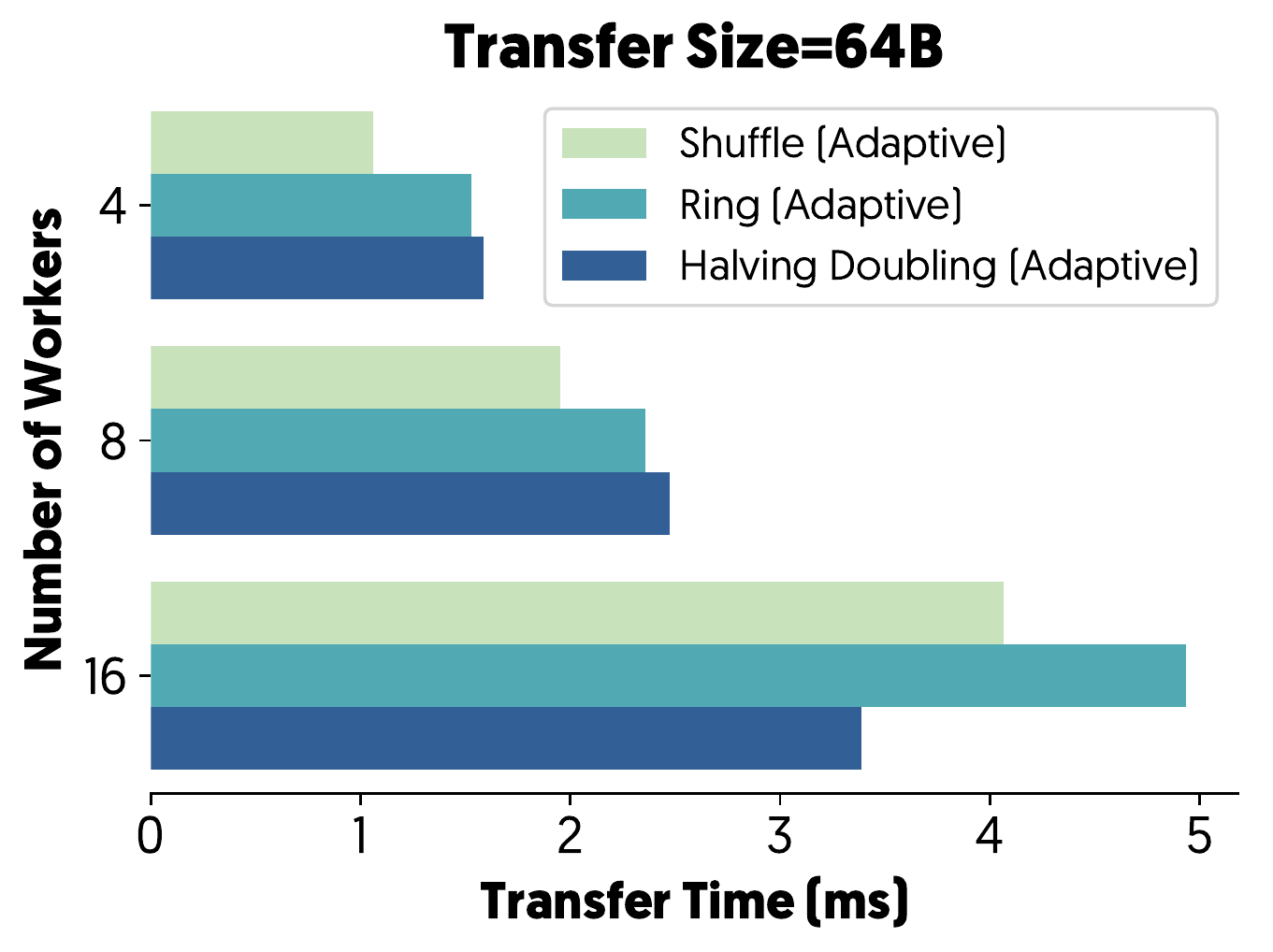}
  \label{fig:ar-implementation-small}
  \caption{Small Transfers}
\end{subfigure}
\begin{subfigure}[t]{0.42\textwidth}
  \centering
  \includegraphics[width=0.8\textwidth]{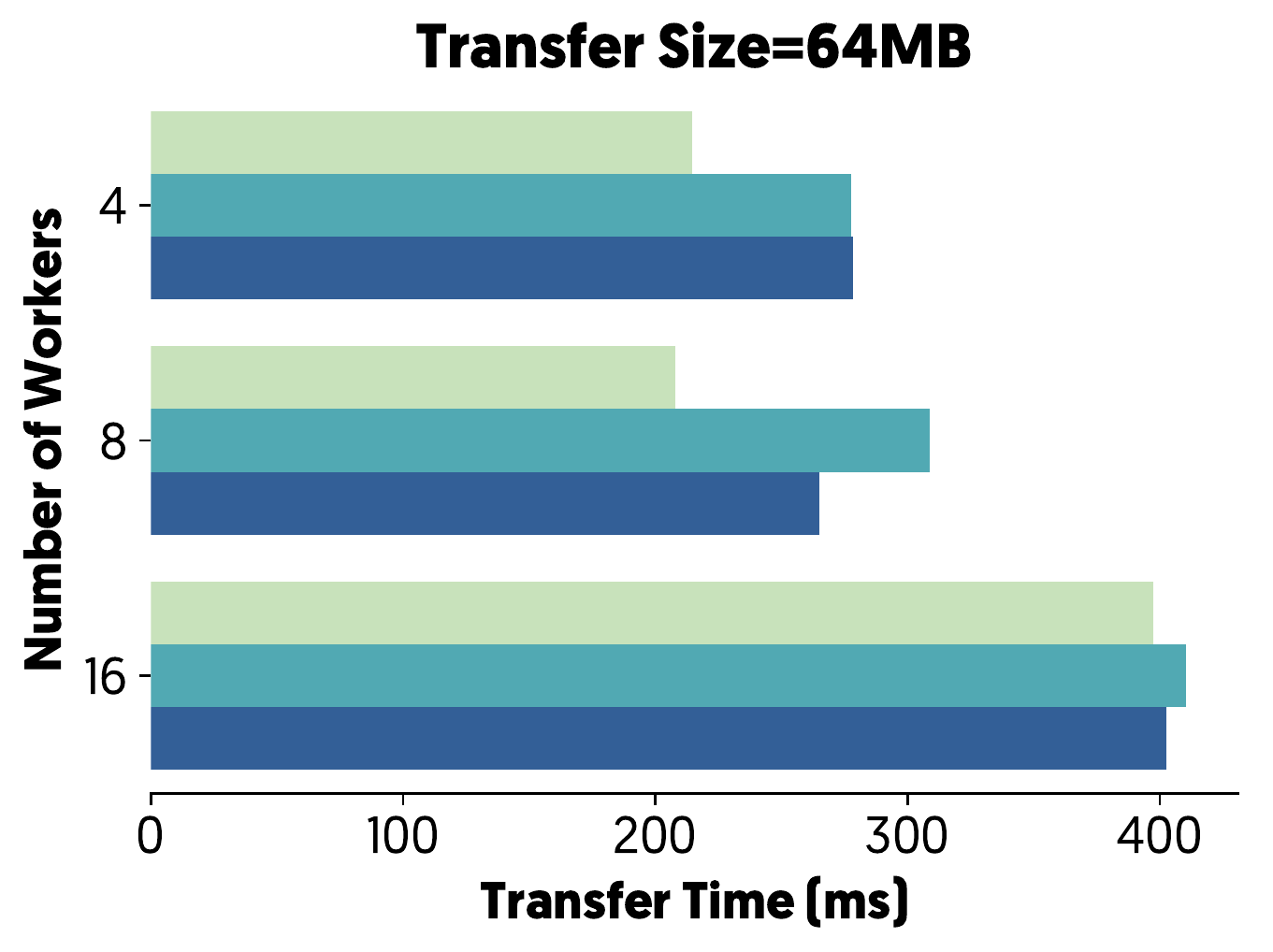}
  \label{fig:ar-implementation-large}
  \caption{Large Transfers}
\end{subfigure}
\caption{Performance Comparison of Different Collective Transfer Implementation in Caramel. Lower is better.}
\label{fig:ar-implementation}
\vspace{-2mm}
\end{figure}

All results until the previous subsection are based on shuffle mode of decentralized aggregation. However, \sysname optimizations are applicable to all decentralized aggregation schemes. In FIgure~\ref{fig:ar-implementation}, we show the iteration time with two other decentralized schemes: ring and halving-doubling at two transfer sizes representating small and large transfers. Shuffle has the highest performance benefit with less number of workers, hence we showed results for this scheme. As the number of workers increase, halving-doubling has better performance. The choice of the best aggregation scheme depends on the number of workers, network bandwidth available, etc.

\parab{In summary}, we have shown that \sysname offers the following performance benefits:
\begin{itemize}[leftmargin=*]
    \itemsep0em
    \item \sysname improves iteration time by up to $3.62\times$ and GPU utilization by up to $3.5\times$ compared with Horovod in $5$ popular DNNs.
    \item Optimizations in \sysname reduces communication cost and improves the communication/computation overlap.
    \item Small parameter batching and adaptive depth allows \sysname to choose the optimal chunk size for transmission with minimal overhead.
    \item \sysname is the first system implementing decentrilized aggregation to support network transfers during the forward pass of computation, thereby increasing overlap significantly. 
\end{itemize}

\section{Discussion}
In this section, we discuss limitations of \sysname and avenues for future work.

\vspace{-1mm}
\parab{Dynamic and variable models:} \sysname cannot accurately predict the timing of dataflow models with Dynamic control flow~\cite{yu2018dynamic} or models with highly variable input sizes (e.g. DeepSpeech2 \cite{deepspeech2}) since our model relies on the iterative nature of the DNN training. In such environments, inaccurate prediction can lead to higher iteration time.



\vspace{-1mm}
\parab{Extending network optimization to multiple GPUs:} \sysname focuses on optimizing network transfers over the cloud network. Our implementation does not rely on Nvidia's NCCL or other GPU-to-GPU libraries to aggregate the data on a single machine. In future, \sysname can be extended with additional optimization for network aggregation between multiple GPUs within a single machine. 

\vspace{-1mm}
\parab{Alternative implementations:} Our implementation currently generates an In-Graph dataflow model, where the dataflow at all workers is represented in a single large DAG and later partitioned. The size of this graph grows as the number of workers increases, which may increase the TensorFlow processing time at large graph sizes. Note that we have not hit this limit with the current models. In contrast, Horovod uses a Between-Graph dataflow model, where each worker's version of dataflow model is generated separately. Since none of the optimizations in \sysname is dependent on the type of the dataflow model, \sysname components may also be implemented as a "between-graph".

\vspace{-1mm}
\parab{Extending to other frameworks: } \sysname is currently implemented over TensorFlow. However, the optimizations are independent of the choice of framework, and can be adapted to other systems (similar to porting Horovod from TensorFlow to PyTorch \cite{paszke2017pytorch}). 

\section{Related Work}
Several solutions have been proposed for reducing iteration time through network acceleration in distributed DNN training. The first category focuses on modifying the machine learning algorithm with the objective of optimizing network tranfers~\cite{alistarh2017qsgd, wen2017terngrad, 203269}. \sysname does not change the DNN, it only adds additional dependencies in the dataflow DAG without altering the underlying logic. The second class of solutions decreases network overhead by reducing the precision of parameters~\cite{vanhoucke2011improving,courbariaux2015binaryconnect,gupta2015deep}. \sysname does not change the parameters of the DNN. The third approach is to optimize the aggregation pattern for accelerated DNN training~\cite{goyal2017accurate, cho2017powerai,  deepspeech2, DBLP:journals/corr/abs-1709-05011, DBLP:journals/corr/abs-1711-04325, 203269}. \sysname belongs to this category. However, prior solutions for improving communication/computation overlap~\cite{arnold2016ddl,cui2016geeps,203269} developed for earlier layer-by-layer systems where the model is sequential cannot be adapted to modern DAG-based systems. \sysname algorithms are not related to these prior solutions.

Solutions for improving communication/computation overlap in Parameter Server (PS) based systems cannot optimize Collective communication (AR) due to significant differences in execution model. PS has 3 steps: “Push” gradients to PS, “Update” parameters on PS, and “Pull” parameters to workers. Poseidon~\cite{203269}, P3~\cite{p3}, and TicTac~\cite{tictac} overlap Pull, Update, and Push across different parameters at the same time in PS-based aggregation. AllReduce (which \sysname tackles) has only 2 steps: “collective reduce” of gradients followed by “Update” parameters at each worker (Fig 1). More importantly, similar techniques are used differently in Caramel and past work. E.g., \sysname splits transfers to overlap "time on wire" with kernel context switching and aggregation op within a single transfer. P3 splits a transfer to overlap Push and Pull of subparts. \sysname transfers all subparts in parallel while P3 transfers sequentially. 



Kylix~\cite{kylix} proposed the use of allreduce primitives (such as recursive halving-doubling used by \sysname) in commodity clusters primarily for big data processing systems such as Hadoop and PowerGraph. It leverages sparsity of data to optimize network transfers. While \sysname relies on the same primitives, we implement additional optimizations tailored to the TensorFlow framework. Moreover, \sysname chooses \textit{when} to do the aggregation and on \textit{what data size} based on the model and network characteristics. Another work~\cite{hpc-rel1} that optimizes allreduce for machine learning frameworks  is tailored for HPC environment with high speeds and not suitable for the cloud environment (InfiniBand is $54$+Gbps and Azure cloud environment provides the highest cloud bandwidth of $10$Gbps.)

 Horovod~\cite{horovod}, built atop an earlier work \cite{baidu-allreduce}, uses decentralized aggregation pattern with model-replica training jobs similar to \sysname. Horovod also adds communication ops to dataflow DAG of TensorFlow. However, it redirects the communication to MPI or NCCL allreduce implementations with limited optimizations on transfer. In contrast with Horovod, \sysname involves significant optimization for overlap improvement and communication cost reduction using fine-grained scheduling and batching. The large performance benefits of \sysname over Horovod is due to these model- and network-aware optimizations.
 
 ByteScheduler~\cite{bytescheduler}, a generic scheduler for both PS and AllReduce with network-only optimizations, has limitations for AllReduce workloads. It needs custom implementation for every accelerator and network fabric. Currently, it only supports NVIDIA GPUs but not CPU/TPU~\cite{byteschCode}. \sysname works with all hardware supported by TensorFlow out of the box. ByteScheduler also requires out-of-DAG implementation of parameter optimization and only supports SGD, Adam and RMSprop currently~\cite{byteschCode}. \sysname supports all TensorFlow optimizers and auxilliary services such as checkpointing without any modification. Random execution order of transfers which could cause deadlock and underutilized network and pipelining of parameters are other problems in AllReduce that only \sysname tackles.

 More importantly, prior work in this space require changes to the underlying framework: P3~\cite{p3} modified KVServer in MXNet, TicTac~\cite{tictac} modified WorkerService in TensorFlow, ByteScheduler~\cite{bytescheduler} uses out-of-DAG scheduler. \sysname works with vanilla TensorFlow without any changes to the underlying framework.

\section{Conclusion}
Iteration time in distributed DNN training in cloud environment is often bottlenecked by network transfers. In this paper, we develop \sysname to accelerate DNN training through network transfer optimizations. \sysname identifies the appropriate aggregation pattern for a given network environment to reduce the communication cost. The communication/computation overlap is improved with model- and network-aware optimizations. High performance gains achieved by decentralized aggregation patterns in \sysname motivates further research in decentralized aggregation mechanisms tailored for cloud environments.

{\small 
\bibliographystyle{sysml2019}
\interlinepenalty=10000
\bibliography{caramel.bib}
}

\end{document}